\documentclass[a4paper]{article}
\usepackage{lmodern}
\usepackage{amssymb,amsmath}
\usepackage{ifxetex,ifluatex}
\usepackage{fixltx2e} 
\ifnum 0\ifxetex 1\fi\ifluatex 1\fi=0 
  \usepackage[T1]{fontenc}
  \usepackage[utf8]{inputenc}
\else 
  \ifxetex
    \usepackage{mathspec}
    \usepackage{xltxtra,xunicode}
  \else
    \usepackage{fontspec}
  \fi
  \defaultfontfeatures{Mapping=tex-text,Scale=MatchLowercase}
  
\fi
\IfFileExists{upquote.sty}{\usepackage{upquote}}{}
\IfFileExists{microtype.sty}{%
\usepackage{microtype}
\UseMicrotypeSet[protrusion]{basicmath} 
}{}
\usepackage[margin=1in]{geometry}
\usepackage{longtable,booktabs}
\usepackage{graphicx}
\makeatletter
\def\maxwidth{\ifdim\Gin@nat@width>\linewidth\linewidth\else\Gin@nat@width\fi}
\def\maxheight{\ifdim\Gin@nat@height>\textheight\textheight\else\Gin@nat@height\fi}
\makeatother
\setkeys{Gin}{width=\maxwidth,height=\maxheight,keepaspectratio}
\ifxetex
  \usepackage[setpagesize=false, 
              unicode=false, 
              xetex]{hyperref}
\else
  \usepackage[unicode=true]{hyperref}
\fi
\hypersetup{breaklinks=true,
            bookmarks=true,
            pdfauthor={},
            pdftitle={},
            colorlinks=true,
            citecolor=blue,
            urlcolor=blue,
            linkcolor=magenta,
            pdfborder={0 0 0}}
\let\rmarkdownfootnote\footnote%
\def\footnote{\protect\rmarkdownfootnote}

\usepackage{titling}

  \title{}
  \pretitle{\vspace{\droptitle}}
  \posttitle{}
  \author{}
  \preauthor{}\postauthor{}
  \date{}
  \predate{}\postdate{}

\usepackage{setspace}
\usepackage{lineno}

\begin{document}

\maketitle

\section{Assessment of the potential impacts of plant traits across
environments by combining global sensitivity analysis and dynamic
modeling in
wheat}\label{assessment-of-the-potential-impacts-of-plant-traits-across-environments-by-combining-global-sensitivity-analysis-and-dynamic-modeling-in-wheat}

\subsection{Authors}\label{authors}

Pierre Casadebaig (1)*, Bangyou Zheng (2), Scott Chapman (3), Neil Huth
(4), Robert Faivre (5), Karine Chenu (6)\\(1) INRA, UMR1248 AGIR, 31326
Castanet-Tolosan, France\\(2) CSIRO Agriculture Flagship, Queensland
Bioscience Precinct, 306 Carmody Road, St.~Lucia, QLD 4067,
Australia\\(3) Queensland Alliance for Agriculture and Food Innovation
(QAAFI), The University of Queensland, St Lucia, QLD 4350,
Australia\\(4) CSIRO Agriculture Flagship, 203 Tor Street, Toowoomba,
QLD 4350, Australia\\(5) INRA, UR875 MIAT, 31326 Castanet-Tolosan,
France\\(6) Queensland Alliance for Agriculture and Food Innovation
(QAAFI), The University of Queensland, 203 Tor Street, Toowoomba, QLD
4350, Australia\\(*) Corresponding author

\subsection{Keywords}\label{keywords}

genotype-environment interactions; ideotype; breeding; sensitivity
analysis; modeling; crop model; APSIM; wheat; Australia; drought

\subsection{Abstract}\label{abstract}

\begin{quote}
A crop can be viewed as a complex system with outputs (e.g.~yield) that
are affected by inputs of genetic, physiology, pedo-climatic and
management information. Application of numerical methods for model
exploration assist in evaluating the major most influential inputs,
providing the simulation model is a credible description of the
biological system. A sensitivity analysis was used to assess the
simulated impact on yield of a suite of traits involved in major
processes of crop growth and development, and to evaluate how the
simulated value of such traits varies across environments and in
relation to other traits (which can be interpreted as a virtual change
in genetic background). The study focused on wheat in Australia, with an
emphasis on adaptation to low rainfall conditions. A large set of traits
(90) was evaluated in a wide target population of environments (4 sites
x 125 years), management practices (3 sowing dates x 2 N fertilization)
and \(CO_2\) (2 levels). The Morris sensitivity analysis method was used
to sample the parameter space and reduce computational requirements,
while maintaining a realistic representation of the targeted trait x
environment x management landscape (\(\sim\) 82 million individual
simulations in total). The patterns of parameter x environment x
management interactions were investigated for the most influential
parameters, considering a potential genetic range of +/- 20\% compared
to a reference.\\Main (i.e.~linear) and interaction (i.e.~non-linear and
interaction) sensitivity indices calculated for most of APSIM-Wheat
parameters allowed the identifcation of 42 parameters substantially
impacting yield in most target environments. Among these, a subset of
parameters related to phenology, resource acquisition, resource use
efficiency and biomass allocation were identified as potential
candidates for crop (and model) improvement.
\end{quote}

\newpage

\subsection{Introduction}\label{introduction}

Progress in plant breeding is limited by the ability to predict plant
phenotype based on its genotype, especially for complex traits such as
yield. Suitably constructed process-based models provide a mean to
reduce this gap in particular by dissecting the complexity of the
genotype-environment interactions and by simulating expected impacts in
various environmental conditions {[}1--3{]}, including consideration of
future climates {[}4,5{]}.\\From a modeling point of view, crops are
complex systems arising from interactions among genetic determinants,
physiological processes, pedo-climatic factors and management practices.
The combination of these elements, which are either chosen (cultivar and
management) or given (soil and climate) in any sown crop, generates
greatly variable stress patterns {[}6,7{]} and results in high genotype
(G) x environment (E) x management (M) interactions. A number of such
interactions has been reported in the literature {[}8,9{]}, and sources
of yield variation, especially in rainfed systems, commonly arise
primarily from the genotype x environment (GxE) interactions, rather
than the genotype (G), i.e.~GxE \textgreater{} G as observed for field
pea in Canada {[}10{]}, sunflower in Argentina {[}11{]}, sorghum in
Australia {[}12{]}, wheat in north-east Australia {[}13{]} and globally
{[}14{]} and maize in Midwestern US states {[}15,16{]} Modeling
approaches have been developed to better understand GxExM interactions
and attempt to take advantage of genetic and environmental resources
more efficiently. For example, Hammer et al. {[}17{]} show that the
multi-year risk of crop failure for farms within a given sorghum region
can be reduced by the adoption of better combinations of GxM (``local
G'' and ``local M'') compared to use of the combination of ``global G''
and ``global M'' that would be adopted if using the entire sorghum
production area.

Process-based crop models are useful tools to integrate scientific
knowledge and simulate varietal or management impacts on productivity in
the target population of environments (TPE), i.e.~the set of
environments to which newly bred varieties need to be adapted
{[}18,19{]}. Hence, the predictive capability of crop models is used to
explore the complex GxExM landscape and assists breeding programs to
take advantage of genetic and environmental resources more efficiently
{[}2,20,21{]}. While such models are based on mathematical equations
translating biological processes in relation to crop growth and
development, their parameters can be controlled to mimic effects of
genotypic variability and explore the GxExM landscape using
\emph{virtual genotypes} {[}22,23{]}. Numerical exploration of crop
models for the target population of environments thus allows exploration
of the entire GxExM landscape, assuming that the crop simulation model
gives a credible description of the biological system.

To be relevant, exploration of the GxExM landscape has to be applied to
environments and management practices related to targeted production
systems. A recent study characterized the drought environment of rainfed
wheat for the Australian target population of environments {[}7{]}, an
interesting target given that Australia is the fourth wheat exporter
worldwide and that Australian wheat crops have to adapt to a high
variability (spatial and inter-annual) in drought patterns, which
strongly impedes crop breeding {[}9,13,24{]} The Australian wheatbelt
extends ca. 13 million ha (Australian Bureau of Statistics, 2013) and
has soils ranging from shallow sandy to deep clay soils and include
temperate, Mediterranean and subtropical climates {[}25,26{]}. Chenu et
al. {[}7{]} undertook a simulation-based study (60 sites x 5 initial
soil moisture x 5 sowing dates) to capture the variability in
environmental and management conditions of this TPE. To study genotypic
variation in such a TPE raises computational challenges if variations in
multiple plant traits with high granularity (resolution) are desired,
i.e.~requiring the simulation of many levels of small increment for each
of the factors explored.

The APSIM (www.apsim.info) Wheat model {[}27--29{]} is used to simulate
crop performance as a function of plant traits, pedo-climatic
variability and management practices. This model has been extensively
used and tested across Australia {[}6,27,28,30{]}. Numerical experiments
with crop models allow exploration of large GxExM landscape. However,
sampling the GxExM landscape using a factorial design with as few as six
levels for each parameter of the APSIM-Wheat model in the Australian TPE
considered in this study would require to perform
9.72x10\textsuperscript{73} simulations. Such an approach would require
absurdly high computing resource and could be considered as partly
wasteful given that it considers all parameters including those of
minimal importance. An alternative is to apply a numerical method
designed to more efficiently explore complex landscapes. For instance,
global sensitivity analysis allows investigation of how the uncertainty
in the output of a model can be apportioned to different sources of
uncertainty in the model input {[}31,32{]}.

Few computational studies have used sensitivity analysis to address
cropping problems, e.g assessing the impact of phenology and management
on sugarcane yield in various environments {[}33{]}, the influence of
geometrical and topological traits on light interception efficiency of
apple trees {[}34{]} and the impact of physiological traits on wheat
grain yield and protein concentration in Europe {[}35{]}. Recently, Zhao
at al. {[}36{]} performed a sensitivity analysis on the APSIM-Wheat
model with a focus on a narrow set of cultivar-specific traits (10
parameters) with the aim to improve an incoming calibration step.

The aims of this paper were (i) to assess the impact of a suite of
physiological traits on yield for Australian rain-fed wheat crops and
(ii) to evaluate how the value of such traits varies across environments
and in relation to other traits. A large set of traits (\(103\)) were
evaluated in APSIM-Wheat for a wide population of environments related
to four representative locations {[}7,24{]} and 125 years of historical
records of weather data (Fig. 1). In addition to this representative set
of 500 environment conditions, simulations were performed for three
sowing dates, three levels of nitrogen fertilization and two levels of
\(CO_2\) (i.e.~9000 conditions in total) to assess the effects of
management and \(CO_2\) factors. We used a global sensitivity analysis
to determine the effects of all traits on yield for all the conditions
studied (i.e.~each site x year x management combination). Traits found
to have substantial and frequent impacts on yield were further studied
through variance analysis to investigate the influence of
environmental-factors and their impact on integrated traits such as
plant leaf area, biomass production, and grain size and number.

\subsection{Material and methods}\label{material-and-methods}

\subsubsection{Overview}\label{overview}

A global sensitivity analysis was applied on the APSIM-Wheat crop model
to identify potential candidate traits for yield improvement in a large
population of environments. Figure 1. describes this workflow, showing
how the ``genetic diversity'' was considered, sampled and screened
\emph{in silico}. In summary, from 516 parameters of the APSIM-Wheat
model, 90 independent parameters that could be considered as ``component
traits'' were selected to reflect a potential genetic variability. Each
of the 90 component traits was assumed to vary in a \(\pm\) 20\% range
around the value for the reference cultivar \emph{Hartog}. The number of
considered traits prevented the use of a factorial design, and so the
Morris method {[}37,38{]} was used to sample the total parameter space
(90 traits, 6 levels, 100 reps; i.e.~9100 ``genotypes''). Simulations
for those genotypes were performed with APSIM-Wheat (Version 7.5) for
(1) 4 locations (Fig. 2) and 125 years (from 1889 to 2013, Table 1) to
test the impact of component traits in the TPE and (2) for 3 sowing
dates (i.e.~early, TPE-level and late) , 3 levels of nitrogen (i.e.~low,
TPE-level and high fertilization) and 2 levels of \(CO_2\) (380 and 555
ppm to represent \(CO_2\) level in 2010 and 2050) to test trait impact
in other environmental conditions related to farmer management practices
and future climates. The impact of the 90 component traits were
considered for 8 output variables (``integrated traits'', Table 2)
related to phenology (flowering and maturity dates), leaf development
(Leaf Area Index at flowering), biomass production (at maturity), and
grains (grain number, size, protein and yield; Fig. 4-6). Overall 42
component traits were identified as ``influential'' (i.e.~main average
impact on yield greater than 20 kg ha-1; Fig. 4) and considered as
potential candidates to improve yield in the TPE. They were analyzed in
more detail with a variance analysis (Fig. 7). Several interesting
traits related to phenology, resource acquisition, resource use
efficiency and biomass allocation were studied in more detail as their
impact could be related to specific environmental factors (Fig. 8). A
more complete description of the workflow and analysis is given below.

\includegraphics{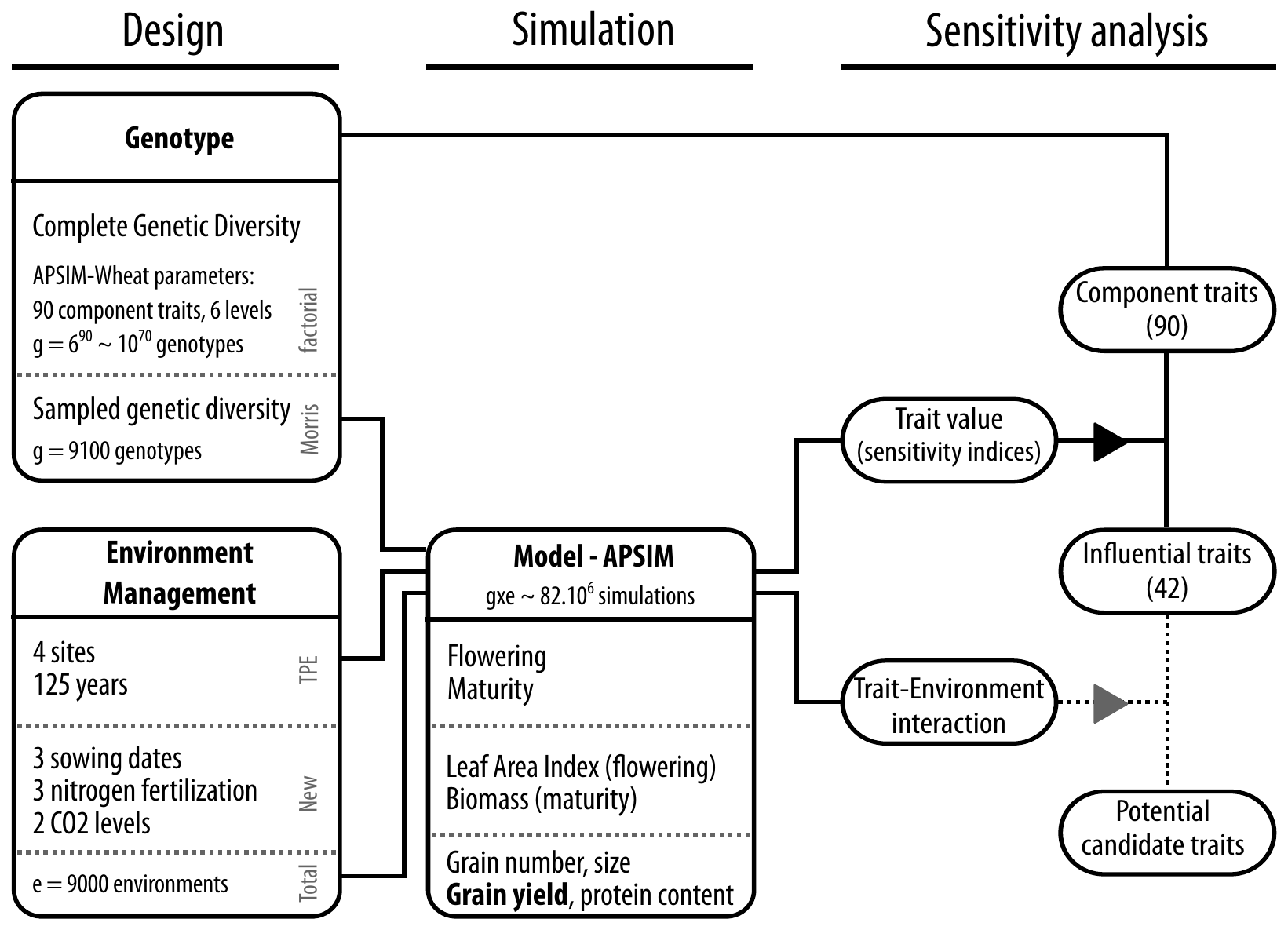}

\textbf{Figure 1. Framework of crop model simulation and the sensitivity
approach used to assess the potential impact of plant traits} \emph{A
global sensitivity analysis was applied on the APSIM-Wheat crop model to
identify potential candidate traits for yield improvement in a large
population of environments. This workflow presents how the ``genetic
diversity'' was considered, sampled and screened in silico. In summary,
90 independent APSIM-Wheat parameters considered as ``component traits''
were selected to reflect a potential genetic variability. Each of the 90
component traits was assumed to vary in a \(\pm\) 20\% range around the
value for the reference cultivar ``Hartog''" and the Morris method
{[}37,38{]} was used to sample the total parameter space (90 traits, 6
levels, 100 reps; i.e.~9100 ``genotypes''). Simulations for those
genotypes were performed with APSIM-Wheat (Version 7.5). The impact of
the 90 component traits were considered for 8 output variables
(``integrated traits'', Table 2). The impact on crop yield allowed to
screen component traits for influential traits (n=42) in the target
population of environments while a study on trait x environment
interactions was used to explore their variability across environments.}

\subsubsection{Simulations and sensitivity
analysis}\label{simulations-and-sensitivity-analysis}

A global sensitivity analysis was performed on parameters of the crop
model APSIM-Wheat version 7.5 {[}28,29{]} to assess their impact on
yield in the Australian wheatbelt (Fig. 1-2). Five main steps were
followed: (1) listing the input APSIM-Wheat parameters (input factors)
to be included in the analysis, (2) setting the variation range for each
factor, (3) sampling the parameter space with the Morris method, (4)
simulating the virtual experiment with APSIM-Wheat and (5) computing the
sensitivity indices to assess the impact of each factor singly (main
effect) or in combination (interaction).

\paragraph{1. Defining input factors}\label{defining-input-factors}

As for most crop models, APSIM-Wheat has parameters (Table S1) that
specify quantitative effect of processes related directly or indirectly
to crop growth and development {[}27--29,39{]}. Those parameters are
typically either single values or arrays of paired vectors (Table S1;
Fig. S1), in which case one vector relates to the piloting a state
variable (x; e.g.~stage values) and the second one corresponds to the
considered trait (y; e.g.~values of root biomass partitioning for the
different key stages considered). Each defined value, whether it is a
single-value parameter or a point in an array can be considered as a
parameter; in which case, APSIM-Wheat (v. 7.5) has 516 parameters
{[}29{]}.

Not all parameters were considered when assessing the impact of plant
traits on crop performance as (1) parameters representing soil physics
and general physical constants were not considered, (2) parameters
deliberately set to have no impact on wheat crops were not considered
(e.g.~multiplicative scalars which are set to 1.0 by default in the
released version of APSIM-Wheat) and (3) values in vectors (parameter
arrays) were considered as dependent parameters, counting one parameter
for the whole ``function''. This reduced the number of parameters to 103
(62 single values and 41 functions). In addition, some parameters were
grouped {[}38{]} to avoid aberrant situations and computational errors
(e.g new min thresholds being greater than new max thresholds). In
total, 20 parameters (annotated with * in Table S1) were grouped into 7
``meta-parameters'' that govern their variation (e.g.~nitrogen demand,
leaf expansion processes). Overall, 90 parameters
(\(p = 103 - 20 + 7 = 90\)) were considered in the sensitivity analysis.

\paragraph{2. Setting the variation
range}\label{setting-the-variation-range}

The range of parameter values is biologically constrained by the genetic
diversity existing in wheat. However, most crop models have typically
been designed to only simulate major differences among cultivars
(e.g.~phenology), as their primary aim has been to address crop
management problems. As a result, crop models such as APSIM-Wheat only
have a few parameters that are by default considered as
cultivar-dependent, while all the other parameters are assumed to be
constant for the species. Given the lack of knowledge related to the
range of the genetic variability existing for most of the model
parameters, a fixed range of 40\% variation for all parameters was
tested in the sensibility analysis. Where possible, equal variation
around the nominal value (\(\pm\) 20 \%) was considered, but for
hard-bounded parameters (e.g scalars comprised between 0 and 1) the 40\%
variation was considered below (or above) the nominal value. Nominal
values were considered for the reference cultivar \emph{Hartog} and
scaled using two consecutive rules: (1) direct scaling of the single
value, or of all the \emph{y} vector for function parameters
(e.g.~proportion of biomass partitioned to the roots at different
stages) and (2) scaling only one single point in the \emph{x} or
\emph{y} vector when this improved the biological meaning
(e.g.~threshold of leaf-expansion sensitivity to water deficit). Figure
S1 illustrates the shape and variation range for function parameters
studied in this sensitivity analysis.

\paragraph{3. Sampling of the parameter space and experimental
design}\label{sampling-of-the-parameter-space-and-experimental-design}

We used the Morris method {[}37{]} as implemented by Campagnolo et al.
{[}38{]} to sample the parameter space and compute sensitivity indices.
The method consists in a discretization of the input space for each
factor (n = 6 levels), then performing a given number of one-at-a-time
(OAT) design (r = 100). The OAT designs were randomly chosen in the
input space, and the variation direction was also random. The repetition
100 times of these steps allowed the estimation of elementary effects
for each input factor. The implementation in the \emph{sensitivity} R
package used a space-filling optimization of the design {[}38{]}.
Parameter design was normalized to account for the different magnitudes
in input factors (parameters expressed in different units).\\Considering
the total number of input factors and the sampling conditions, the total
size of parameter design was \(90+1 \times 100 = 9100\), where each
sample (i.e.~set of parameter values) can be interpreted as a virtual
genotype (i.e.~9100 in total). The numerical sampling of the parameter
space can be viewed as an exploration of virtual genotype materials
where there is no restriction in the combination of traits considered
(i.e.~no genetic linkage or epistasis).

\paragraph{4. Crop simulations}\label{crop-simulations}

The parameter-sampling design consisting of 9100 virtual genotypes was
used to simulate the crop performance for these genotypes in different
environmental conditions and thus evaluate mean parameter impact and
parameter x environment interactions.

APSIM-Wheat simulations were first done for the target population of
environments (i.e control conditions, Table 1) defined by 4 sites
(Emerald, Narrabri, Yanco and Merredin; Fig. 2, Table 1) and 125 years
(1889-2013) of climatic data (\(4 \times 125 = 500\) environments). Crop
management in these simulations (Table 1) was chosen to mimic local
farming practices {[}7{]}. Additional simulations were performed for 3
sowing dates 21/04; 15/05; 07/06), 3 nitrogen fertilization levels (low:
50\% of TPE-level, TPE-level and high fertilization: TPE-level plus 50
kg.ha\textsuperscript{-1}) and 2 \(CO_2\) levels (TPE-level of 380 ppm
and 555 ppm to represent \(CO_2\) level in 2010 and 2050) to explore the
impact of parameters in contrasting N and \(CO_2\) conditions.\\Nitrogen
fertilization rules followed APSIM decision model: at sowing, nitrogen
was applied as nitrate in Merredin and as urea in the rest of the
wheatbelt. In Yanco, fertilisation at ``end of tillering'' stage only
occurred if cumulative rainfall since sowing was greater than 100 mm,
and fertilisation at ``mid-stem elongation'' stage only occurred if
plant available water was greater than 60\% of the PAWC. At Merredin,
fertilisation at ``mid-stem elongation'' only occurred if plant
available water was greater than 60 mm.\\Overall, 9000
(\(3 \times 3 \times 2 \times 500 = 9000\)) environmental conditions
were tested, and 81.9 million of crops (\(9100 \times 9000\)) were
simulated on the CSIRO distributed computing cluster which can sustain a
peak throughput of approximately 8000 simultaneous processes {[}40{]}.
Parameter impacts were tested on eight output variables from APSIM
(Table 2): number of days from sowing to flowering and from sowing to
maturity, leaf area index (LAI), biomass production, the number, size
and protein content of grains and yield.\\The baseline simulations were
performed with the reference cultivar Hartog to estimate environmental
indices (Table 2) and crop performance in each environment. In addition,
the growing environments were characterized in terms of drought
environment types, as described in Chenu et al. {[}7{]}.

\includegraphics{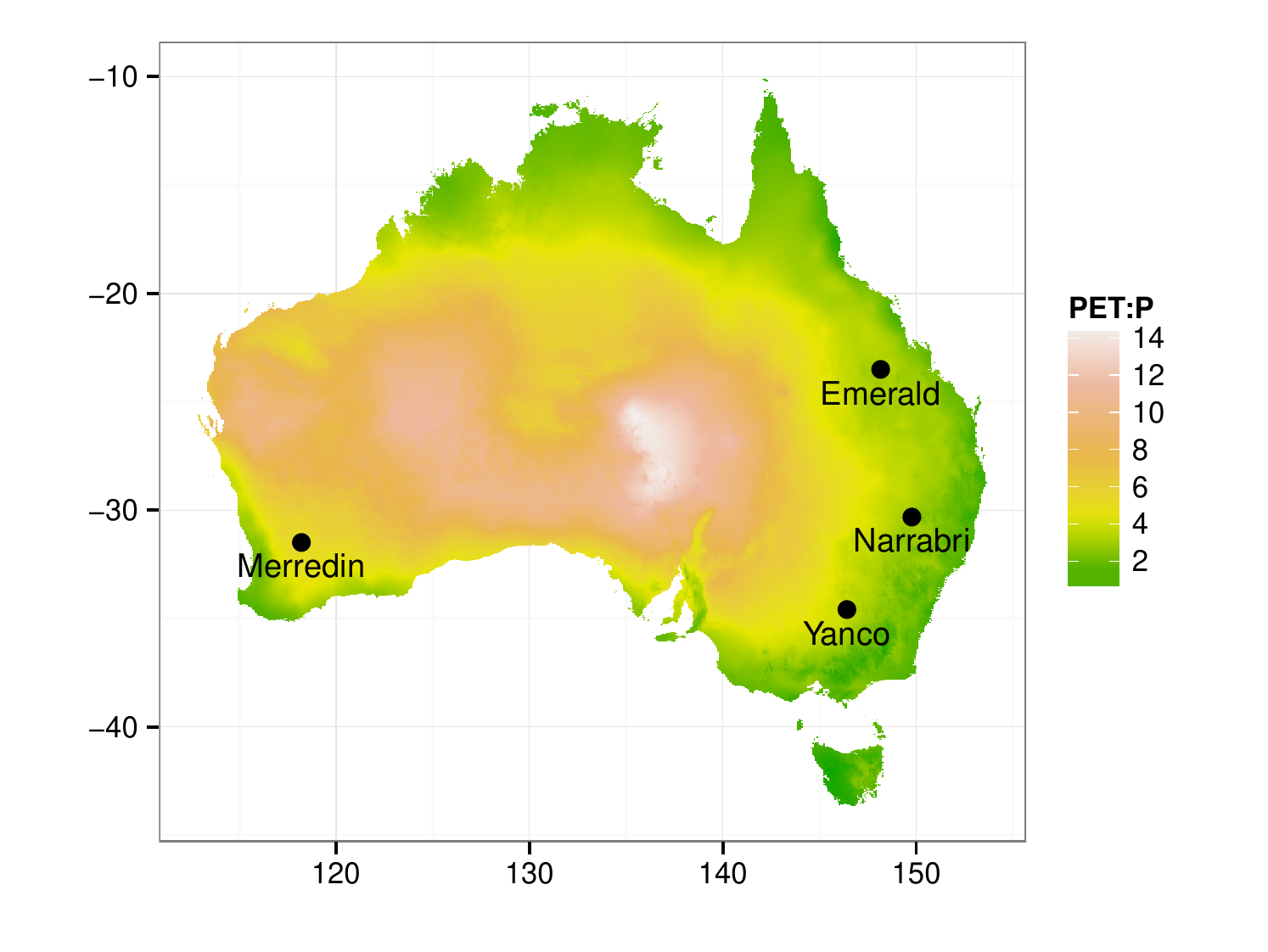}

\textbf{Figure 2. Map of the studied sites and climatic variability in
aridity index.} \emph{The map shows potential evapotranspiration over
precipitation ratio (1 / aridity index, data from Zomer et al.
{[}41{]}), points correspond to locations sampled in the target
population of environments.}

\begin{longtable}[c]{@{}lllll@{}}
\toprule
& Emerald & Narrabri & Yanco & Merredin\tabularnewline
\midrule
\endhead
latitude (degree) & -23.53 & -30.32 & -34.61 & -31.5\tabularnewline
longitude (degree) & 148.16 & 149.78 & 146.42 & 118.22\tabularnewline
rainfall pattern & summer dominant & summer dominant & evenly
distributed & winter dominant\tabularnewline
annual rainfall (mm) & 635 & 650 & 425 & 303\tabularnewline
seasonal rainfall (mm) & 170 & 249 & 228 & 209\tabularnewline
seasonal PET (mm) & 843 & 640.2 & 462.2 & 601.6\tabularnewline
daily mean temperature (celcius) & 18.4 & 13.9 & 11.9 &
13.1\tabularnewline
daily mean radiation (MJ.m\textsuperscript{-2}) & 18.3 & 15.7 & 13.3 &
14.5\tabularnewline
soil type & black vertosol & grey vertosol & brown sodosol & shallow
loamy duplex\tabularnewline
PAWC (mm) & 133.5 & 217.5 & 190.8 & 101.1\tabularnewline
sowing date & 15/05 & 15/05 & 15/05 & 15/05\tabularnewline
sowing PAWC (mm) & 132 & 175 & 99 & 39\tabularnewline
initial nitrogen (kg.ha\textsuperscript{-1}) & 30 & 30 & 50 &
30\tabularnewline
applied nitrogen (kg.ha\textsuperscript{-1}) & 50/0/0 & 130/0/0 &
40/40/40 & 20/20/30\tabularnewline
\bottomrule
\end{longtable}

\textbf{Table 1. Characteristics of the locations, soils and management
representing the target population of environments.} \emph{Plant
available water capacity (PAWC) is indicated for each soil, as well at
the level of initial soil water used in the simulations (median of plant
available water at sowing which was estimated from {[}7{]}). Applied
nitrogen dose are indicated by ``a/b/c'': respectively, the
fertilization applied at sowing (a), at the stage ``end of tillering''
(b) and at the stage ``mid-stem elongation'' (c). Annual and seasonal
(1-May to 1-Nov) climatic data were considered for 1889-2013.}

\paragraph{5. Computation of sensitivity
indices}\label{computation-of-sensitivity-indices}

Sensitivity indices were computed as statistics of elementary effect,
i.e effect of the factor for each repetition {[}37,38{]}. In this
approach, the main effect (noted \(\mu^*_i\) in Iooss et al. {[}42{]})
is a measure of the influence of the \emph{i}-th input on the output,
and is calculated as the mean of the absolute value of the elementary
effects. The larger \(\mu^*_i\) is, the more the input contributes to
the dispersion of the output. The interaction effect (\(\sigma_i\) in
Iooss et al. {[}42{]}), is a measure of non-linear and/or interaction
effects of the \emph{i}-th input. \(\sigma_i\) is computed as the
standard deviation of the elementary effects. An input with a large
\(\sigma_i\) can be considered as having non-linear effects or being
involved in an interaction with at least another input. We also computed
a standardized sensitivity index to be able to compare indices across
different output variables (as in Fig. 5.) and growing conditions (as in
Fig 8.). In this case, for each growing environment, the model output
variables were standardized (\(x' = \frac{x - mean(x)}{sd(x)}\)) before
computing elementary effects and sensitivity indices.

\begin{longtable}[c]{@{}llll@{}}
\toprule
Type & Variable & Description & Unit\tabularnewline
\midrule
\endhead
Crop & Flowering & Flowering date & day\tabularnewline
Crop & Maturity & Maturity date & day\tabularnewline
Crop & LAI & Leaf area index at flowering & -\tabularnewline
Crop & Grain Size & Dry biomass of an individual grain &
g\tabularnewline
Crop & Grain Number & Grain number & grain\tabularnewline
Crop & Grain Protein & Grain protein content & \%\tabularnewline
Crop & Biomass & Crop aerial dry biomass at harvest & t
ha\textsuperscript{-}1\tabularnewline
Crop & Yield & Crop grain yield at harvest & t
ha\textsuperscript{-1}\tabularnewline
Environment & Water & Average soil water deficit ratio &
-\tabularnewline
Environment & Nitrogen & Average nitrogen stress factor &
-\tabularnewline
\bottomrule
\end{longtable}

\textbf{Table 2. Description of integrated traits (APSIM-Wheat output
variables) and environmental indices included in the analysis.}
\emph{Environment indices were computed for the sowing-harvest period,
for all considered environments. Water-deficit index correspond to the
simulated water supply-demand ratio and relates to the degree to which
the water available to the roots matches the plant water demand {[}7{]}.
Nitrogen stress index relates to the level of nitrogen stress on
photosynthesis. Stress indices are expressed as scalars so that values
range from 0 (low stress) to 1 (high stress).}

\subsubsection{Clustering parameters according to their
impact}\label{clustering-parameters-according-to-their-impact}

All the considered parameters were subdivided into three groups
according to the mean value of their main effect in the TPE (i.e.~mean
of \(\mu^*_i\) across environments): (1) \emph{null impact} group, in
which parameters had no impact on crop yield in any environments (2)
\emph{low impact} group, in which the parameters had an average
\(\mu^*_i\) lower or equal to 0.02 t ha\textsuperscript{-1} and (3)
\emph{impactful} group, in which parameters had an average main effect
on yield that was greater than 0.02 t ha\textsuperscript{-1}. A
hierarchical clustering based on Ward distance was applied to the matrix
of \emph{impactful} parameters and the eight output variables (averaged
across environments) to group these parameters and identify those with
similar patterns of effect on output variables.

\subsubsection{Environment indices and trait x environment
analysis}\label{environment-indices-and-trait-x-environment-analysis}

For parameters identified by the sensitivity analysis as yield-impacting
traits in the TPE, a variance analysis was performed to assess the
effects of environmental factors on the parameter main-impact
variability. Hence, for each trait, a linear model was fitted with
environment-related factors (\(CO_2\), sites, sowing, nitrogen)
considered as fixed effects and with no interaction. The effect of each
environmental factor (\(e\)) on trait impact was estimated by the
proportion of total sum of square (\(\eta^2\)) as \(SS_e/(TSS)\). Note
that both the effect of ``uncontrollable'' environmental factors
(i.e.~climate) and the interactions among factors were pooled in the
residuals.

Finally, we considered the response to the environment of a small subset
of candidate traits and defined several environmental stress indices
(Table 2) to further illustrate the ecophysiological basis of trait x
environment interactions. Using the ASPIM Wheat model, daily computed
indices related to water and nitrogen stresses were averaged for the
duration of the crop cycle. In the model, water-stress is computed as a
function of the soil water extractable by roots (water supply) and
potential crop transpiration (water demand) {[}7{]}. The nitrogen-stress
determined limiting nitrogen level affecting leaf photosynthesis
{[}29{]}. In this study, both indices were set to range from 0
(no-stress) to 1 (extreme stress) to allow comparison between stress
indices.

\subsubsection{Software}\label{software}

All data processing, statistical analysis and graphics were performed
with R 3.1.0 {[}43{]} with additional R packages \emph{dplyr} (data
processing {[}44{]}), \emph{sensitivity} (sensitivity analysis, version
1.10.1 {[}45{]}) and \emph{ggplot2} (visualization {[}46{]}).

\subsection{Results}\label{results}

\paragraph{A target population of environments with contrasting
environmental
conditions.}\label{a-target-population-of-environments-with-contrasting-environmental-conditions.}

Four sites were chosen to capture part of the variability in soil types
and rainfall patterns that are experienced across the dryland wheatbelt
(Table 1; Fig. 1). Simulated yield for 1889-2013 reflected these
differences in environments, with median yield ranging from 1.72 t ha-1
in Emerald to 4.10 t ha-1 at Narrabri (Fig. 3). High inter-annual
variability was also simulated and reflected the broad range of water
deficits and temperature events that Australian wheat experience across
seasons {[}5,7{]}.

\includegraphics{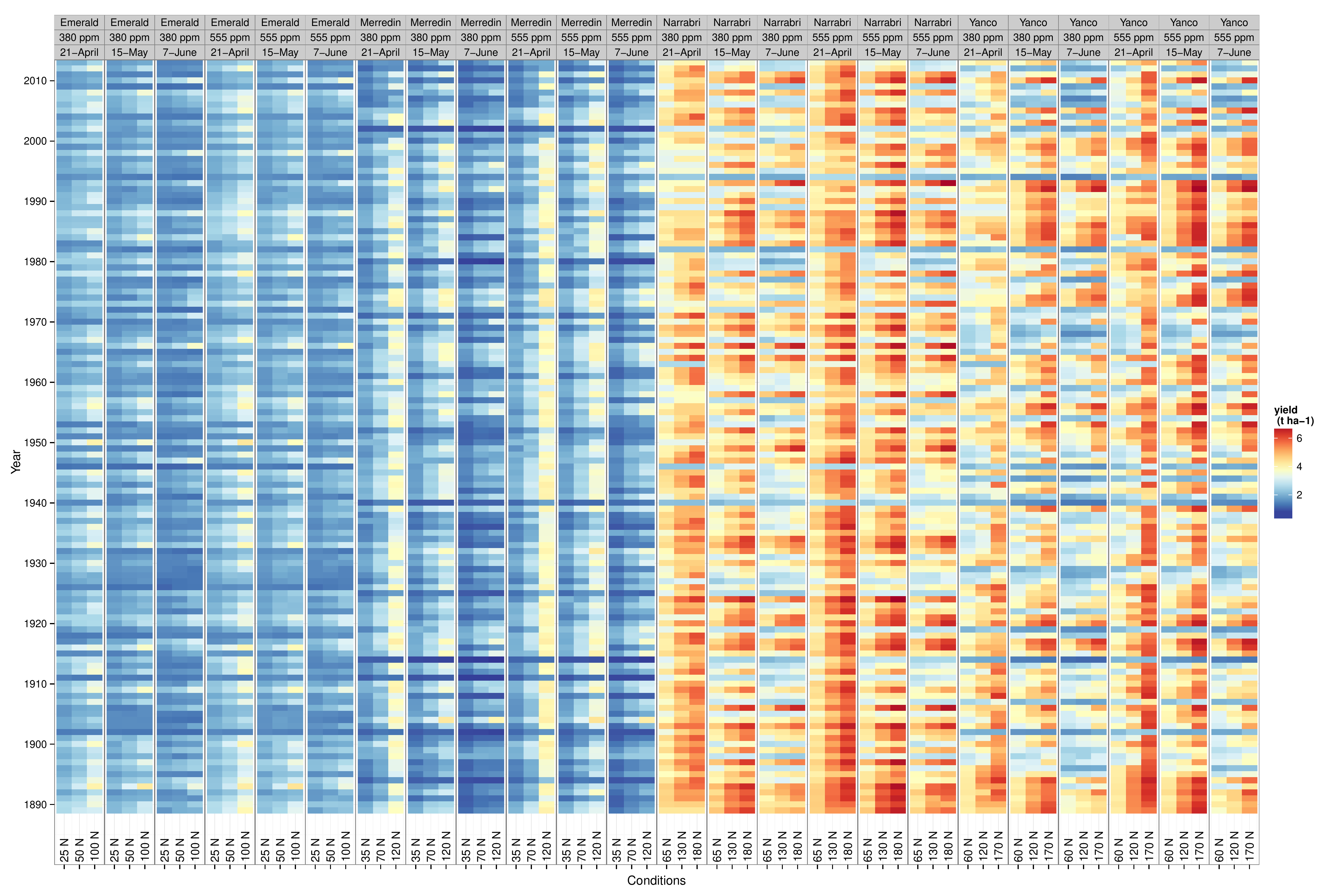}

\textbf{Figure 3. Heatmap of yield response to climate and management
practices in all growing environments studied.} \emph{Simulated yield
for cv. ``Hartog'' is presented for each sites (Emerald, Merredin,
Narrabri, Yanco), \(CO_2\) levels (380 and 555 ppm), sowing dates (21
April, 15 May, 7 June), fertilization (x-axis, potential mineral
nitrogen applied before decision model, in kg ha\textsuperscript{-1})
and climatic years (y-axis) i.e 9000 growing environments in total.}

\paragraph{About a half of the studied traits had little or no impact on
yield in the target population of environments
(TPE).}\label{about-a-half-of-the-studied-traits-had-little-or-no-impact-on-yield-in-the-target-population-of-environments-tpe.}

A global sensitivity analysis was performed to get a general picture of
the effect of APSIM-Wheat parameters on yield response in the TPE. While
the results from the sensitivity analysis strongly depend on the ranges
of variation for the input traits, such ranges are scarcely available
for all the considered traits despite numerous studies and reviews
giving informative indications of partial genetic ranges for some traits
{[}47--50{]}. To perform a broad screen of parameters, the sensitivity
analysis was done with variations of \(\pm\) 20 \% from the reference
value (\emph{Hartog} cultivar) of each parameter (Table S1), except for
some function parameters for which variations were adapted to increase
the biological likelihood of the results (see Fig. S1). Another analysis
was conducted with variation of \(\pm\) 50 \% to test a broader range of
variation, but this led to a high proportion of crop failure, due in
particular to excessive senescence (data not shown).

\includegraphics{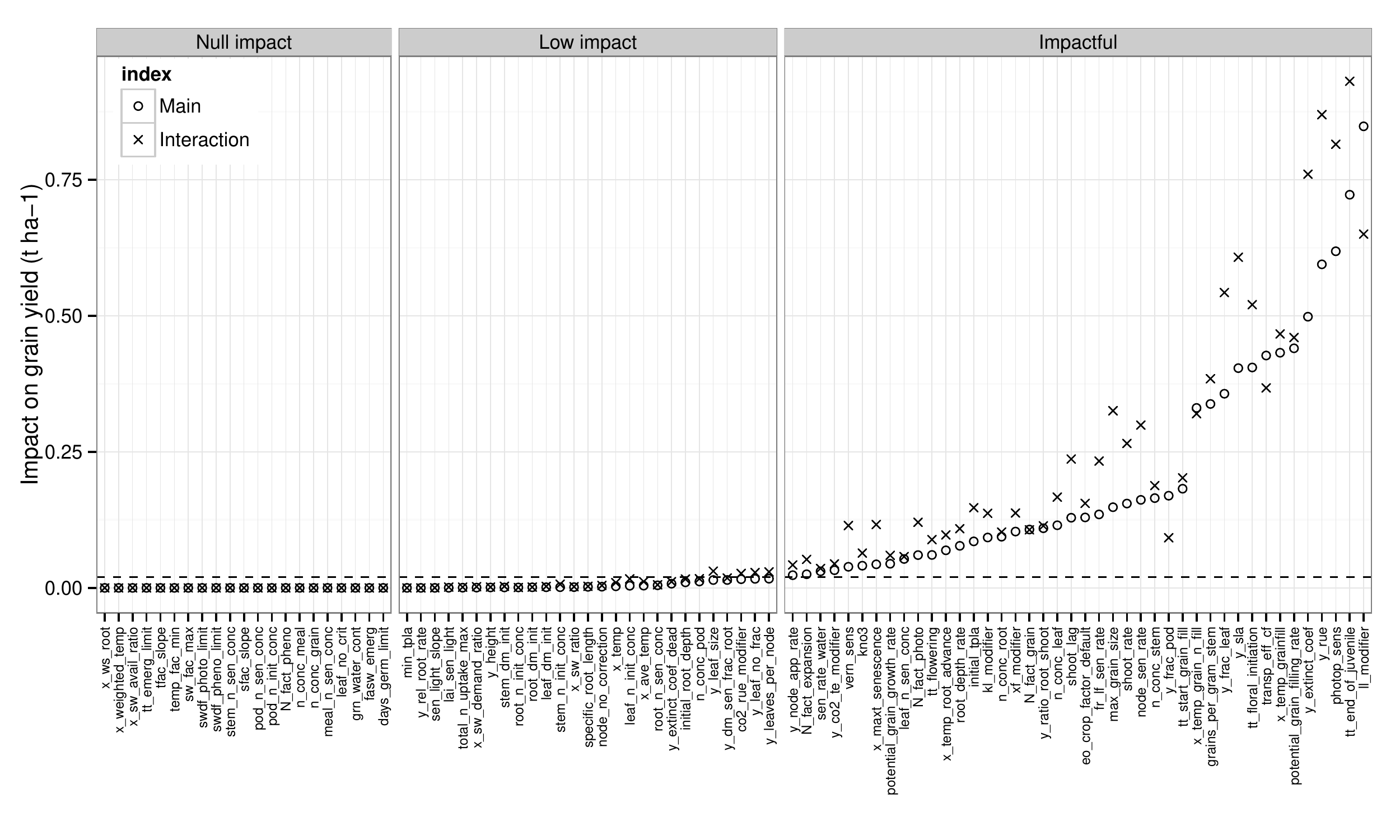}

\textbf{Figure 4. Screening for influent traits in the TPE (control
conditions)}. \emph{Traits were ranked by increasing mean main
sensitivity index and were grouped into three groups (panels): ``null
impact''``,''low impact" and ``impactful'' group. Note that all impacts
are positive, as given by the sensitivity analysis method. A description
of traits is presented in supplementary Table 1. Concerning sensitivity
indexes, the main effect (circle) is an estimation of the linear
influence of the considered trait on grain yield, while the interaction
effect (cross) is an estimation of non-linear and/or interaction
effect(s) of the trait. The horizontal dashed line corresponds to the 20
kg ha\textsuperscript{-1} threshold above which traits are considered as
impactful.}

About half of the studied traits (48/90) were not or only weakly
impacting yield (average effect of less than 20 kg
ha\textsuperscript{-1}) in the TPE (Fig. 4). Among those traits, 21 had
no impact on yield or any other of the studied output variables
(i.e.~flowering, maturity, LAI, biomass, grain number, size and protein)
in any environments. Two options could explain such null impacts: (1)
the parameter corresponding to the trait simply did not have any role in
the model algorithm for wheat (some parameters are only used for other
crops in the APSIM framework) or (2) the traits were influent only in
agricultural conditions other than tested here (e.g the sum of
temperature until emergence failure, \emph{tt\_emerg\_limit}).\\The
other 27 traits showed a weak mean impact on yield (\textless{}= 20 kg
ha-1) in the TPE, often because the conditions required to get a
substantial impact are rarely encountered. This group included traits
that may have been considered as important a priori, such as potential
leaf area (\emph{y\_leaf\_size}) or maximum temperature for thermal-time
accumulation (\emph{x\_temp}).\\Traits with a mean impact on yield of
\textgreater{} 20 kg ha-1 were considered in more detail (42 traits;
Fig. 4). Overall, 29 traits had a mean impact between 20 and 25 kg ha-1,
eight traits had an impact between 25 and 50 kg ha-1, and only five
traits had a mean impact greater than 50 kg ha-1. The five most
impactful traits in terms of both mean and interaction effects
(\(\mu^*_i\) and \(\sigma_i\)) in the tested conditions were: the water
extractability by roots (\emph{ll\_modifier}), the thermal time required
to reach floral initiation (\emph{tt\_end\_of\_juvenile}), the
photoperiod sensitivity (\emph{photop\_sens}), the radiation use
efficiency (\emph{y\_rue}), and the radiation extinction coefficient
(\emph{y\_extinct\_coef}).\\Among the 42 impactful traits, only a few
showed a linear impact on yield, i.e.~their main effect was greater than
their interaction effect, e.g.~the fraction of biomass partitioned to
the spike rachis (\emph{y\_frac\_pod}), the water extractability by
roots (\emph{ll\_modifier}), the wheat coefficient for transpiration
efficiency (\emph{transp\_eff\_cf}) and the temperature effect on grain
demand (\emph{x\_temp\_grain\_fill}). Most of the impactful traits had a
ratio of interaction:main effect between 1 and 1.8, denoting either a
large non-linear effect or an effect largely influenced by other traits.
Traits such as senescence-related traits and grain potential biomass
(\emph{max\_grain\_size}) had higher ratio (\textgreater{} 1.8).

\paragraph{Several traits had a strong impact on physiological processes
related to phenology, biomass and grain
production.}\label{several-traits-had-a-strong-impact-on-physiological-processes-related-to-phenology-biomass-and-grain-production.}

To better understand the effects of plant traits in the TPE, the 42
influential component traits were clustered based on their main effect
on eight integrated traits related to phenology, leaf area, and nitrogen
and carbon accumulation and partitioning (Fig. 5). Component traits were
mainly clustered in three groups (dashed line in Fig. 5): lesser
influential traits, traits that strongly impacted all outputs, and
traits that strongly impacted a subset of integrated traits.

\includegraphics{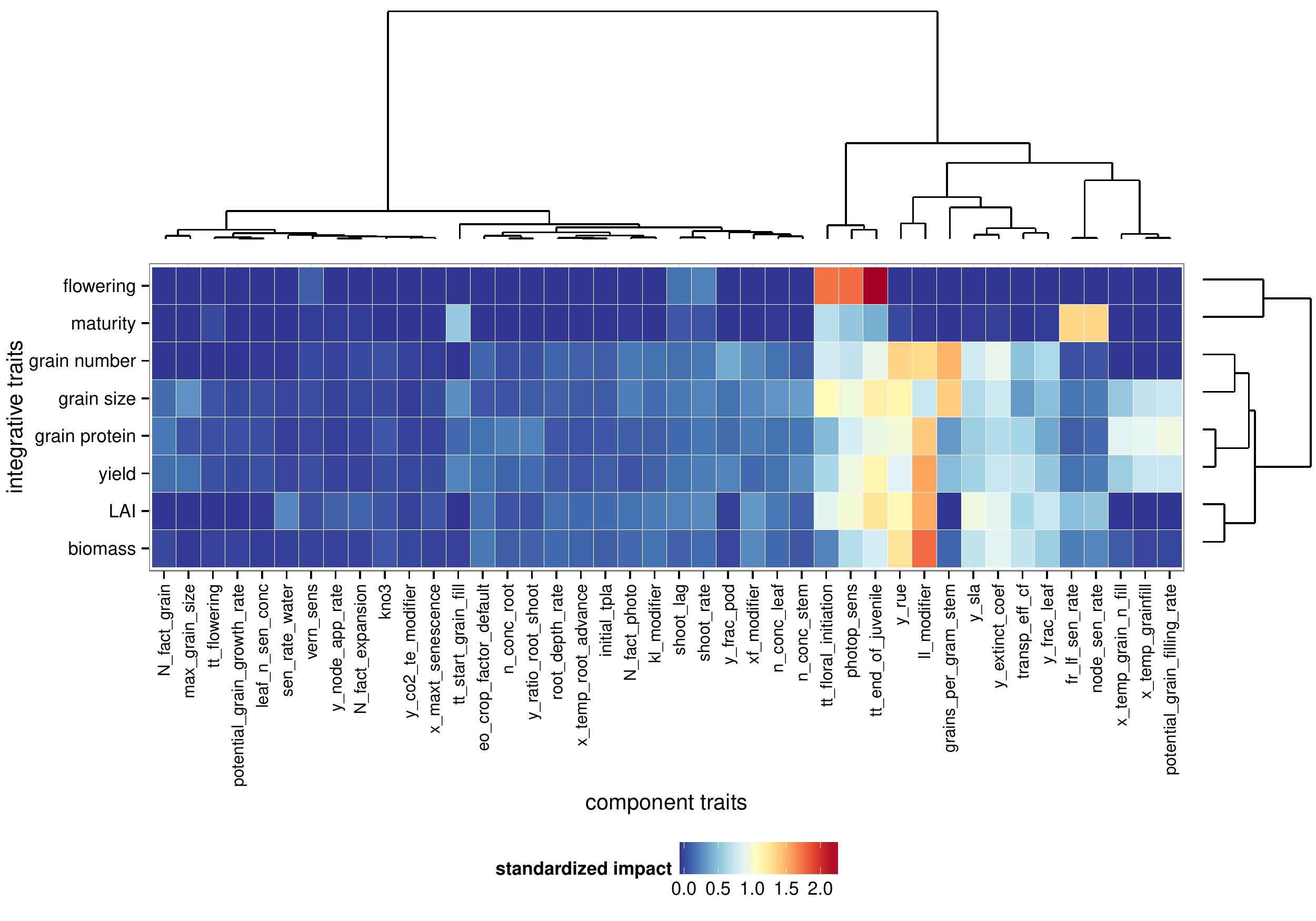}

\textbf{Figure 5. Overview of APSIM-Wheat sensitivity to trait
modification.} \emph{The heatmap shows the impact (positive in the
Morris method) of selected component traits (model inputs, x-axis)
modification on integrated traits (model outputs, y-axis). Component
traits (top dendrogram) and integrated traits (right dendrogram) were
ordered with hierarchical clustering based on the similarities among
impacts. Trait impact was standardized to be comparable across
integrated traits (model output variables).}

Overall, crop phenology (flowering and maturity time) was mostly
affected by six component traits (thermal time from emergence to floral
initiation, from floral initiation to flowering and to a lesser extent
from flowering to the beginning of grain filling; photoperiod
sensitivity and two leaf senescence traits), while the remaining traits
had little to no impact. Traits affecting grain-filling
(\emph{x\_temp\_grain\_n\_filling}, \emph{x\_temp\_grainfill},
\emph{potential\_grain\_filling\_rate}) were clustered together, and had
a high impact on grain size, grain protein and yield. On the other hand,
about another 10 traits were found to substantially impact leaf area,
biomass and grain production. As may be expected, the water
extractability (\emph{ll\_modifier}), which affects the maximum amount
of soil water that can be extracted, impacted traits such as LAI at
flowering, biomass at maturity, grain number and yield. The trait
\emph{grains\_per\_gram\_stem} which relates to the potential of the
crop to set grains based on its carbon status (proportional to stem
weight at flowering), affected grain number but had a relatively little
impact on yield given trade-offs on grain size.\\Globally, the impact
pathway of traits on physiological processes reflected the sub-component
of the crop model where parameters were involved.

\paragraph{Impacts of influential traits were strongly dependent on
environmental and management
conditions.}\label{impacts-of-influential-traits-were-strongly-dependent-on-environmental-and-management-conditions.}

The variability of trait impacts arose from high trait x environment
interactions (Fig. 6), i.e.~the modification of a trait did not result
in the same change in output trait depending on the growing conditions.
Main yield impacts of individual component traits ranged from 0.02 t
ha-1 (screening threshold) to 2.87 t ha-1 (potential radiation use
efficiency, \emph{y\_rue}, under high nitrogen conditions).

\includegraphics{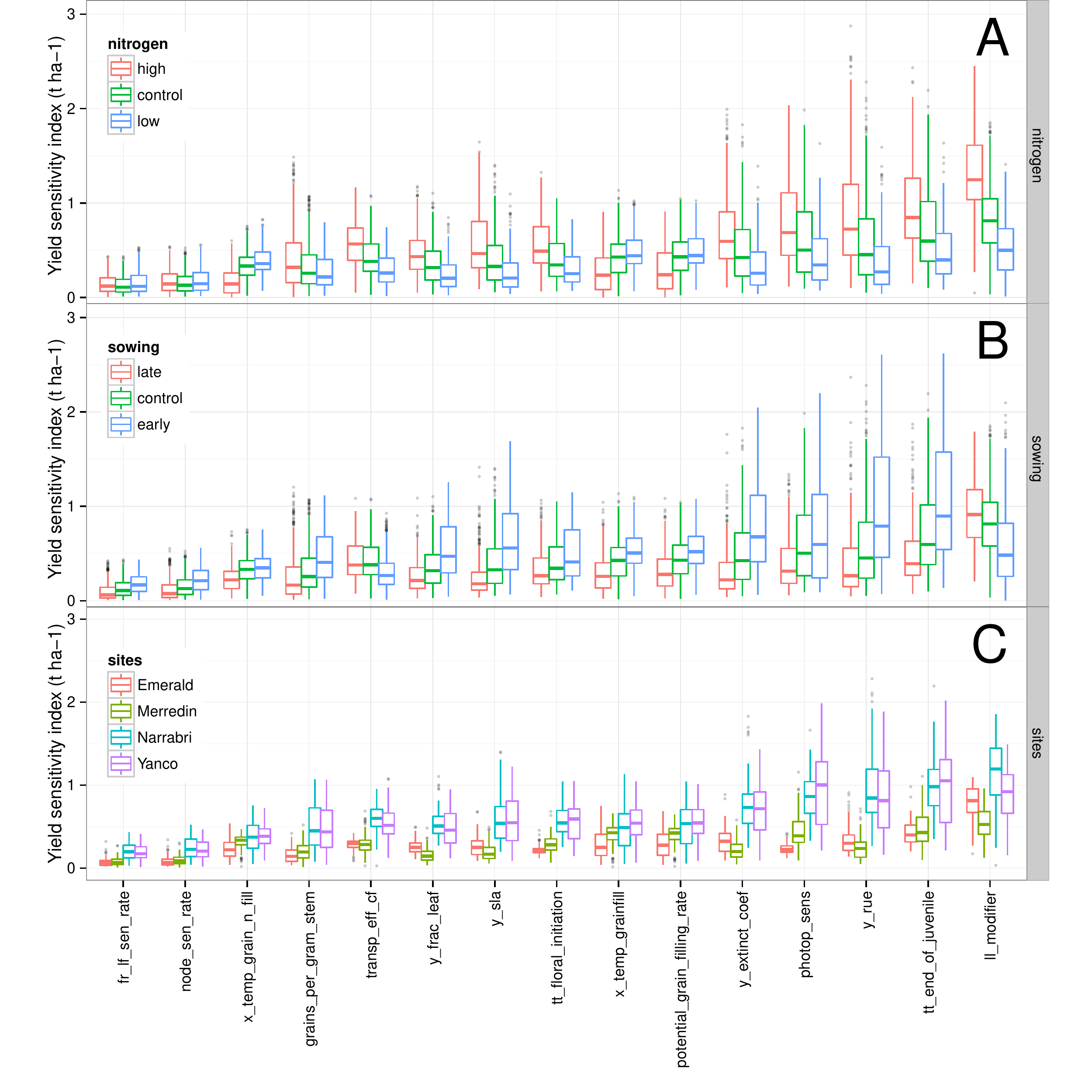}

\textbf{Figure 6. Yield sensitivity to a variation of selected impactful
trait} \emph{Trait main impacts were calculated from a sensitivity
analysis and are presented for different nitrogen treatments (A), sowing
dates (B) and sites (C) and in the TPE (control conditions) unless
mentioned (i.e.~high/low nitrogen, early/late sowing).}

Most traits had a larger yield impact when management practices and
climatic conditions were ``non-limiting'', e.g.~high fertilization, high
soil water holding capacity (Yanco, Narrabri) and early sowing
(i.e.~long cropping season). By contrast, response traits (e.g.
\emph{x\_temp\_grain\_fill}, \emph{transp\_eff\_cf}) impacted yield in
more extensive conditions (e.g.~low nitrogen). For instance, water
extractability by roots (\emph{ll\_modifier}) had more impact for
late-sown than for early-sown crops, as such crops are more prone to
drought.

\paragraph{Identification of influential traits with low dependence to
climate
uncertainty.}\label{identification-of-influential-traits-with-low-dependence-to-climate-uncertainty.}

The variance of trait impacts on yield across the 9000 studied
environments was partitioned for each studied traits into four
controllable environmental factors (\emph{site}, \emph{sowing date},
\emph{nitrogen fertilization} and \emph{\(CO_2\) level}) and one
uncertainty-related factor (\emph{residuals}) that aggregated the factor
\emph{year}, the interaction among ``controllable'' factors and the
residuals (Fig. 7). Despite the coarseness of the approach and the fact
that trait main impacts were only considered as absolute value (no
distinction between negative and positive impact on yield), traits with
both a strong mean impact and an impact variability that mainly depends
on ``controllable'' factors would potentially be easier for
consideration for breeding.

\includegraphics{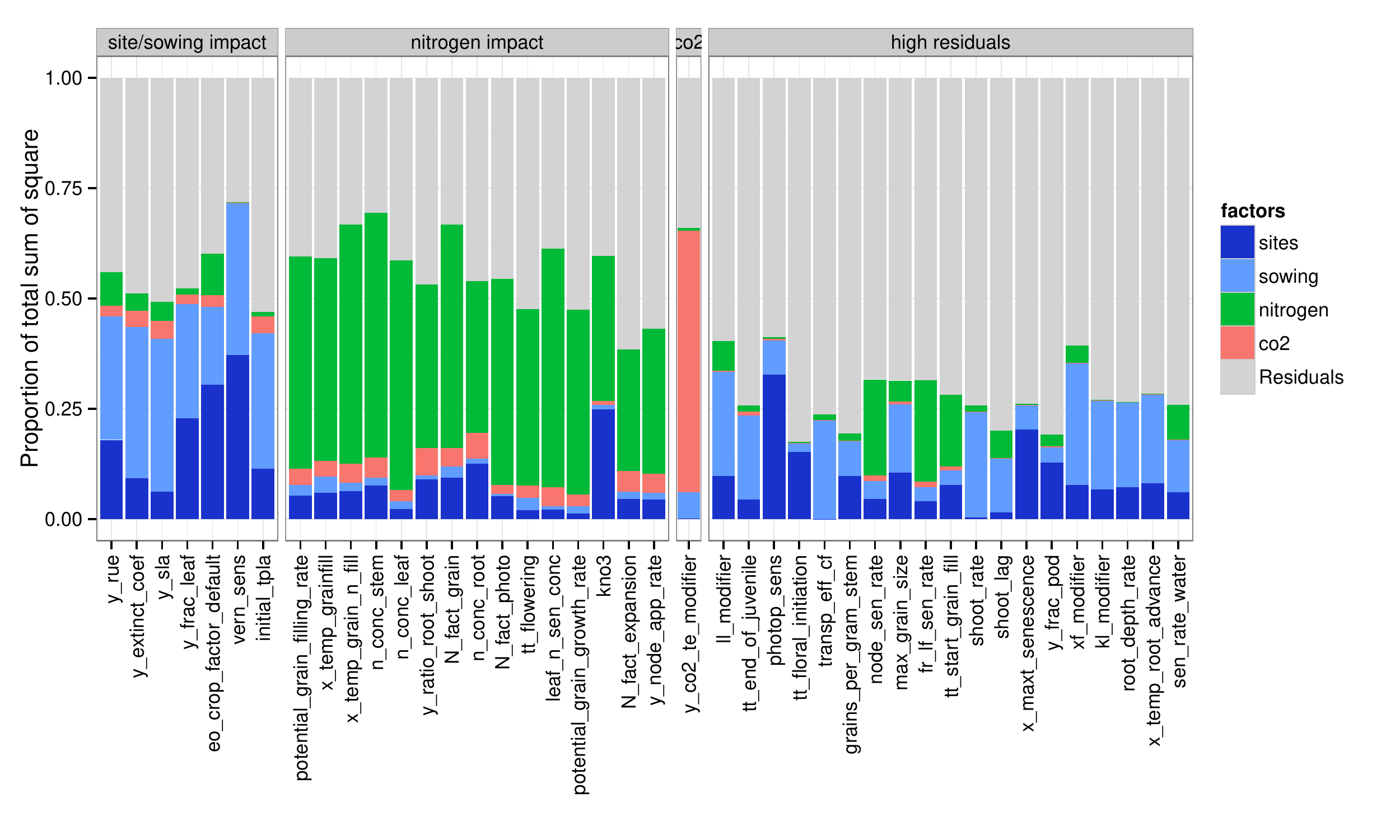}

\textbf{Figure 7. Variance components of trait main impact for major
environmental factors.} \emph{For each influential trait, the proportion
of variance explained by environmental factors (site, nitrogen
fertilization, sowing date and \(CO_2\) level) was calculated in an
ANOVA on simulated yield for crops in the 9000 studied growing
conditions. Traits were clustered in groups based on the proportion of
explained variance by environmental factors (horizontal panels). Cluster
identified corresponded to traits mainly impacted by site and sowing
date (first panel), nitrogen fertilization (second panel), CO2 (third
panel) and traits having a high residual component (fourth panel).}

Traits were \emph{a posteriori} clustered in four groups (horizontal
panels in Fig. 7), which can be described as: (1) site/sowing impact,
which may be related to water or temperature driven processes, (2)
nitrogen impact, (3) \(CO_2\) impact and (4) high residuals
(uncertainty). Traits in the site/sowing, nitrogen and \(CO_2\) groups
displayed both high and relatively stable main impact on yield. The
nitrogen-impact group included all studied traits related to grain
filling, indicating that modifications of such traits could reliably
impact yield providing adequate nitrogen fertilization. On the other
hand, the site/sowing-driven group included traits such as the potential
radiation use efficiency (\emph{y\_rue}), the light extinction
coefficient (\emph{y\_extinct\_coef}) and the potential leaf surface
area (\emph{y\_sla}), which may be linked to the available water
resources or thermal regime (e.g.~short/long crop cycle). Traits in the
\emph{high residuals} group were influential but not stable, meaning
that a modification of such traits did not yield the same return
depending on years and/or due to interaction with other traits.
Phenology-related traits (\emph{tt\_end\_of\_juvenile},
\emph{tt\_floral\_initiation}) and water extractability by roots
(\emph{ll\_modifier}) displayed such behavior, indicating that impact
was likely linked to the level of environmental resources available
(water or temperature, in this case), which is expected in these types
of environments.\\This variance analysis also highlighted expected trait
x environment interactions. For instance, a high \(CO_2\) concentration
triggered the impact of the \(CO_2\) response on transpiration
efficiency (\emph{y\_co2\_te\_modifier}). Note that the effect on
radiation use efficiency (\emph{co2\_rue\_modifier}) was not identified
as influential in the TPE (i.e.~when no change in \(CO_2\); Fig. 4) and
was thus not included in the further analysis. Also, photoperiodic and
vernalization sensitivities (\emph{photop\_sens}, \emph{vern\_sens}) had
contrasting effect across sites and sowing dates. These results are
consistent with field observations.

\paragraph{Trait impacts were related to the availability of
environmental
resources.}\label{trait-impacts-were-related-to-the-availability-of-environmental-resources.}

Strong interactions were identified between environmental factors and
trait impact on yield (Fig. 8) for several traits involved in plant
development (\emph{tt\_end\_of\_juvenile}), resource acquisition
(\emph{ll\_modifier}), biomass production (\emph{y\_rue}) and biomass
allocation (\emph{potential\_grain\_filling\_rate}) processes. Computed
seasonal stress indices for water and nitrogen (see caption of Fig. 8)
were used to highlight these dependencies between environmental stress
and the impact resulting from a trait modification.

\includegraphics{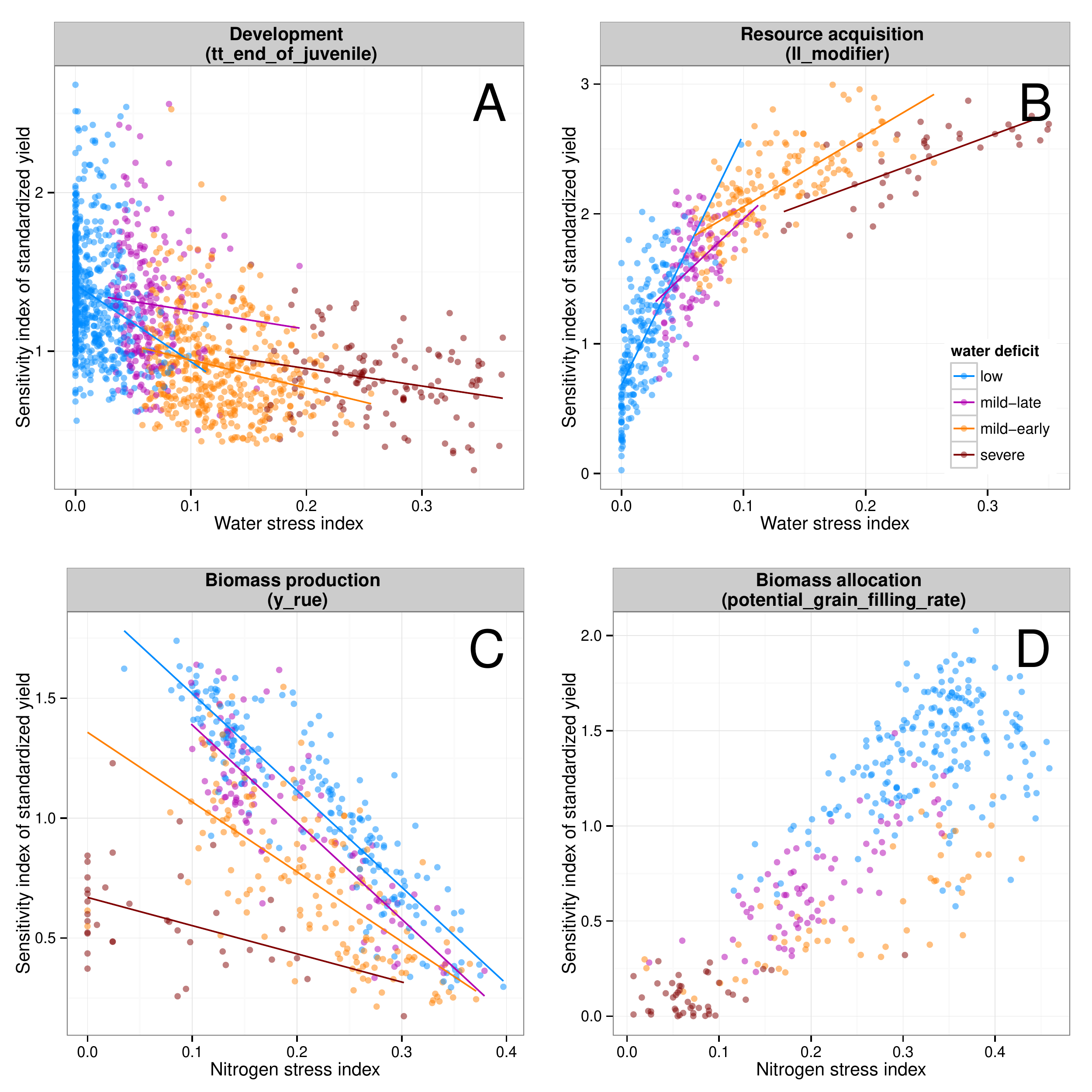}

\textbf{Figure 8. Sensitivity index of standardized yield for selected
component traits involved in crop development (A), resource acquisition
(B), biomass production (C) and biomass allocation (D) relative to
seasonal water- or nitrogen-stress indices.} \emph{Yield impact was
assessed for the thermal time required to reach floral initiation
(tt\_end\_of\_juvenile), the water extractability by roots
(ll\_modifier), the radiation use efficiency (y\_rue), and biomass
allocation to grains (potential\_grain\_filling\_rate). As sensitivity
indices are computed independently for each condition (combinations of
sites x year x management), a standardized sensitivity index was used to
allow comparison of indices across environments. In this case, simulated
yield was standardized (\(x' = \frac{x - mean(x)}{sd(x)}\)) within each
of the 9000 environment conditions before computing elementary effects
and sensitivity indices (which are always positive in Morris method).
The water-stress index {[}7{]} indicates the degree to which the soil
water extractable by roots (water supply) is able to match the potential
crop transpiration (water demand). The nitrogen-stress index is a factor
computed by APSIM that determines limiting N level affecting leaf
photosynthesis {[}29{]}. Both indexes ranged from 0 (no-stress) to 1
(extreme stress).\\Data are presented for representative drought-pattern
environment types (colors), namely ``low'' (ET1) with stress-free or
short-term water-deficits; ``mild-late'' (ET2) with mild water shortage
mainly occurring during grain filling; ``mild-early'' (ET3) with severe
water stress starting during the vegetative stage and relieved during
mid-grain filling; and ``severe'' (ET4) with water deficit from early
stages throughout the grain-filling periods {[}7{]}. Lines represent
linear regressions fitted by environment types.}

Modifications in phenology (\emph{tt\_end\_of\_juvenile}) impacted yield
the most in wet environments (stress index near zero), when yield
potentials were the greatest (Fig. 8A). Nevertheless, this trait had
substantial impacts in all environments, including the most severely
water limited. Change in water extractability by roots
(\emph{ll\_modifier}) also responded to water deficit (Fig. 8B) with
maximum impacts in severe water deficits. Impacts were slightly less
important in mid-early water deficits. They rapidly decreased in less
stressed conditions, but remained substantial. Modifications in
potential photosynthesis (\emph{y\_rue}) had impacts related to both
water and nitrogen availability (Fig. 8C). The relation between impact
and nitrogen availability was linear within each drought environment
type, and the slope of the relation decreased with the severity of the
water deficit (i.e.~the impact response to N was greater in non-limiting
water conditions). Modifications in biomass allocation to grains
(\emph{potential\_grain\_filling\_rate}) led to maximum yield impact in
low water deficit (Fig. 8D) and in severe nitrogen deficits. Yield
impact was increasing with nitrogen deficit but showed a weaker linear
correlation in conditions with severe nitrogen stress (r=0.47).

\subsection{Discussion}\label{discussion}

\paragraph{An \emph{in silico} method to search for potential candidate
traits for
breeding}\label{an-in-silico-method-to-search-for-potential-candidate-traits-for-breeding}

Environmentally-adaptive traits do not scale well from molecular-,
organ- or plant-level to the crop level, particularly when targeting
yield under stressful conditions {[}2,51--53{]}. This difficulty in
demonstrating and estimating the impacts of traits across scales
potentially limits inference of trait value, and is partly responsible
for the non-integration of physiological progress in breeding
programs.\\Here, the problem was approached in the opposite direction
(top-down), to unravel the phenotypic plasticity observed in complex
traits into individual trait contribution at the crop level.
Process-based crop models are designed to integrate physiological
processes and their impact on the local environment (e.g.~soil water
uptake) based on parameters reflecting plant traits (parameterization),
environmental factors and management inputs. As a result, such models
simulate genotype x environment interactions and estimate integrated
traits (e.g.~yield) as emergent properties {[}1,2,22{]}. Here, the
APSIM-Wheat model, which has been widely tested for Australian
conditions was used to weight the impact of numerous plant traits across
the Australian wheatbelt, taking into account climatic variability,
trait x trait interactions and trait x environment interactions.\\While
APSIM-Wheat has over 500 parameters with 103 identified as potentially
varying with genotype, the approach proposed in this paper allowed the
identification of 42 influential traits in the target population of
environments (TPE; Fig. 4). Of these 42 traits, 23 had an impact
relatively stable, meaning that the variance of their impact in was more
explained by ``controllable'' factors (i.e.~site, sowing date, nitrogen
fertilization and CO2 level) and less dependent on climate
uncertainty.\\Overall, the screening phase (sensitivity analysis)
allowed the identification of the most influential traits for yield
(Fig. 4-6); and the searching phase (variance analysis, relation with
specific environmental factors) gave indications as to which traits to
target when considering different types of environments within this
sample of Australian environments, e.g.~high vs low N conditions; Fig. 7
and 8. Such an approach could thus help in estimating trait scalability,
and give a form of return on investment with an estimation of expected
gains from trait modifications. However, additional knowledge is
required when considering the potential value for crop breeding
(e.g.~degree of genotypic variability that may exist for these traits,
trait heritability).

\paragraph{Potential candidate traits for improving yield in the
Australian
wheatbelt}\label{potential-candidate-traits-for-improving-yield-in-the-australian-wheatbelt}

Based on the APSIM-Wheat simulations and a global sensitivity analysis,
traits relative to phenology (\emph{tt\_end\_of\_juvenile},
\emph{photop\_sens}, \emph{tt\_floral\_initiation}), resource
acquisition (water extraction, \emph{ll\_modifier} and light
interception, \emph{y\_extinct\_coef}), resource use efficiency
(\emph{y\_rue}, \emph{transp\_efficiency\_coef}) and biomass allocation
to the grain (\emph{potential\_grain\_filling\_rate},
\emph{grains\_per\_gram\_stem}) were among the most important traits in
the TPE, assuming a ``genetic'' variation of \(\pm\) 20 \% around trait
value of the reference cultivar \emph{Hartog} (Fig. 4). It is important
to keep in mind that the results of a sensitivity analysis strongly
depend on the chosen range of trait variation, and that the 20 \% trait
variation used in this study under-estimated existing variations for
some traits (e.g. \emph{vern\_sens}) while it may have over-estimated
unknown variations in others. However, assuming that the APSIM model
behaves relatively linearly (interaction:main effect ratio of 1-1.5 for
most parameters, Fig. 4.), moderately changing this range would not have
a strong incidence on the estimated impacts (main effect) and providing
new interactions do not arise from the extension of the parameter
ranges. Overall, the approach allowed a first screening of a wide range
of traits for which the range of genetic variability is unknown. This
work could be improved in the future through the incorporation of
knowledge on genetic variability of selected traits.

The most important trait in terms of impact on yield was the water
extractability by roots (\emph{ll\_modifier}; Fig. 4, 5), especially in
Narrabri and Yanco, which had heavy deep soils and thus a high
water-holding capacity (Fig. 6). Genotypic variation in water
extractability at depth was observed in root chambers by Manschadi et
al. {[}50{]}, who assessed that this trait could bring about an extra 50
kg ha-1 for every mm of water extracted during the grain filling period,
for crops grown in the north-eastern part of the wheatbelt (i.e.~ability
to extract more water late in the season has a high marginal value in
terminal stress environments). Compared to other root-related traits,
Veyradier et al. {[}56{]} found that this trait was a strong candidate
for breeding purpose in terms of potential impact. Field experiments for
two cultivars with contrasting water extractability at depth also
highlighted the potential of this trait to improve yield in
drought-prone conditions {[}50,57{]}, which agrees with the increasing
yield impact simulated for increasing drought severity (Fig.
8B).\\Several traits involved in wheat development were identified as
playing a major role in crop performance in the TPE (Fig. 4-6). Traits
related to phenology are usually considered as the primary means to
adapt crops to their growing environments {[}58,59{]}. Recently, an
association mapping study {[}60{]} focused on three traits (earliness
\emph{per se}, photoperiod sensitivity and vernalization requirement),
whose corresponding parameters in APSIM-Wheat model
(\emph{tt\_end\_of\_juvenile}, \emph{photop\_sens} and
\emph{vern\_sens}, respectively) were ranked among the most influential
ones in this study (i.e.~average main impacts on yield respectively of
0.72, 0.62 and 0.04 t ha-1), despite the fact that our reference
cultivar (Hartog) has a low vernalization requirement (\emph{vern\_sens}
of 1.5). These three traits were found to vary in the ranges of 515-980
°Cd, 0-4.1 and 0-2.9 respectively for a broad range of Australian
cultivars {[}21{]}, which is wider than the range tested here (444-666
°Cd, 2.4-3.6 and 1.2-1.8), especially for the vernalization requirement.
The relative importance of those traits on yield is expected to change
when changing their range of variation. In particular, \emph{vern\_sens}
is expected to have a greater impact in the TPE, as found by Zhao et al.
{[}36{]} who tested a range of 0-5 for this trait in a similar analysis.
Also, non-surprisingly, these three traits were found to be strongly
dependent on the site and sowing date (Fig. 7) but had a high level of
variations (high residuals in Fig. 7), which is likely related to
interactions with stresses.\\Other traits had a strong impact on yield.
The most important of these include: (1) the potential RUE
(\emph{y\_rue}) which is a major target for current research projects
aiming to improve photosynthesis efficiency {[}61--64{]}, (2) plant
architecture (\emph{y\_extinct\_coef}) which has been of interest to
some breeders (e.g.~durum-wheat CIMMYT) who have selected for erect
wheat genotypes {[}65{]}, and (3) the potential grain filling rate
(\emph{potential\_grain\_filling\_rate}), which may be improved by the
current efforts of breeders and pre-breeders selecting for stay-green
phenotype {[}66--68{]}, cooler canopy temperature {[}59,69,70{]},
greater reserve remobilisation {[}71,72{]} and/or greater spike
photosynthesis {[}73,74{]}.

\paragraph{The importance of properly considering the target population
of
environments}\label{the-importance-of-properly-considering-the-target-population-of-environments}

Depending on the environment/management conditions considered, the
ranking of trait main impacts varied across traits (Fig. 6), thus
highlighting the need to appropriately consider trait effects across the
target populations of environments {[}75{]}. For instance, the
sensitivity to photoperiod (\emph{photop\_sens}) had a small impact in
Emerald but an important impact in Narrabri and Yanco (Fig. 6C). Hence,
while most influential parameters in Zhao et al. {[}36{]} were also
identified in the most influential subset in our study, the
discrepancies in trait impact between these two studies partly rose from
differences in conditions considered (e.g.~sowing dates, fertilization,
plant density). Our study also explored climate change impacts on the 42
influential traits and indicated that traits of most value may change in
the future, as illustrated for the impact of transpiration-efficiency
response to \(CO_2\) (\emph{y\_co2\_te\_modifier}) under different
levels of \(CO_2\) (Fig. 7). Note that other traits such as
radiation-use-efficiency response to \(CO_2\)
(\emph{co2\_rue\_modifier}), which had only minor impact in current
climates (Fig. 4) and were thus not studied in detail, are likely to
have a substantial impact in the future.

\paragraph{The importance of considering trait combinations rather than
single
traits}\label{the-importance-of-considering-trait-combinations-rather-than-single-traits}

Sadras and Richards {[}52{]} argue and illustrate how indirect breeding
methods often fail to improve yield not because yield is complex, but
rather because those methods do not account for the proper levels of
organization, time scales and interactions among traits and with the
environment. Similarly, trait impacts in crops subjected to multiple
stresses (e.g.~nitrogen and water limitation) are rarely considered in
traditional physiological approaches {[}52{]}. Working with an
integrative crop model, we illustrated in this paper how the potential
value of traits, in combination with others and for a specific TPE, can
be assessed \emph{in silico} by testing (1) if the trait is likely to
impact crop performance (e.g.~estimation of main sensitivity index), (2)
if this impact is modified by controllable (management) or
uncontrollable (climate, genotype x environment interactions) factors,
and (3) how the trait impact is distributed among environment-type of
importance for the TPE {[}7{]}.

The systematic presence of interaction effects found with the
sensitivity analysis (Fig. 4) illustrated that trait interactions are
common. Such results highlight the importance of focusing on collections
of traits rather on individual traits {[}76{]}.\\Furthermore, the close
and complex genetic, physiologic, and agronomic links between carbon,
water, and nitrogen transfers within crops are experimentally difficult
to assess while being important. For instance, efficiencies in water-
and nitrogen-use can be either unrelated, positively (synergy) or
negatively (trade-off) related depending on the environment, the
genotype, the level of organization, and the time scale at which such
efficiencies are defined {[}52,77,78{]}.\\Overall, the complexity of
crop systems highlights the potential of using modeling approaches.
Together with genetic criteria (e.g.~availability of genetic
variability, pleiotropy and heritability) and technical criteria (rapid,
cost-effective, and reliable phenotyping), model-based approaches
(assuming the relevance of the process-based model, of the genetic range
tested and of the TPE) could help breeding to improve crop performance
under changing environments {[}2,52{]}.

\paragraph{A tool to overview and improve crop
models}\label{a-tool-to-overview-and-improve-crop-models}

From a modeling point of view, crop models are evolving over time, while
physiological knowledge underlying crop functioning gradually improves.
Model improvements are thus regularly performed with algorithm
modifications being tracked over time. However, the effects of such
modifications on the model-prediction capacity are usually not clearly
documented nor shared among all model users and developers. Hence, with
different developers focusing simultaneously or successively on a model,
there is a high risk of developing increasingly complex and harder to
understand algorithms. Problems caused by this increased complexity may
affect the quality of the model, but may be revealed and addressed by
using exploration methods throughout model-development phases to
visualize the in-progress modeling state. Global sensibility analyses,
as done in this paper can for instance enable developers to quickly
assess changes in model response due to variation in specific processes,
and notice potential problems.\\In this study, we attempted to consider
the maximum proportion of traits utilized in APSIM-Wheat. The use of
function-table parameters in APSIM complicated the estimation of the
total number of values used as parameters and the assessment of
individual parameter impact on output variables. Overall, about half of
the plant-related parameters of APSIM-Wheat had no impact, keeping in
mind that those parameters may be useful for other crops, or other
processes (e.g.~responses to high-temperature or soil minerals). While
using a global sensitivity analysis to identify such parameters may
appear as an excessive method, the computational cost to include all
parameters (with null, low or high impact) was lower than the time and
expertise needed to analyze the source code and manually identify
subsets of parameters, in the case of this complicated crop model.\\In
total, 42 parameters were identified as \emph{impactful}, as they had an
average main impact greater than 20 kg ha-1 in the TPE. However, only 5
parameters had a mean impact greater than 50 kg ha-1. Martre et al.
{[}76{]} proposed physiological reasons to explain such a surprisingly
low number of influential parameters in crop models: (1) number of
trade-off occur with traits often having compensating effects when
scaling up from plant to crop level (e.g.~once canopies are well
established, increasing the leaf surface area may not improve light
interception and thus photosynthesis) and (2) the fact that complex
characters such as grain yield and protein concentration are inherently
determined at the population level rather than at the organ or plant
level {[}79{]}. While model over-parameterization can result from model
development as well as model design, indicators can help to track the
model complexity and performance. In this context, the use of the
exploration methods described here provides an overview of the model
global response to perturbation (e.g.~Fig 3-5).\\Finally, such
sensitivity analysis can help to identify traits most important for
parameter calibration for cultivars {[}36{]}. Such targeted calibration
can later be implemented with either frequentist {[}80{]} or Bayesian
parameter estimation algorithms {[}81{]}.

\subsection{Conclusion}\label{conclusion}

Phenotyping strategies can be improved by better understanding the
yield-trait performance landscapes {[}82{]}. Here, a global sensitivity
analysis was performed on APSIM-Wheat parameters to identify plant
traits with potential interest for breeding in the Australian wheatbelt.
The genotype x environment x management (GxExM) landscape was explored
for the target population of environments (TPE), with strategic sampling
of APSIM parameters varying for \(\pm\) 20 \% around the reference
values of Hartog. Main (i.e.~linear) and interaction (i.e.~non-linear
and interaction) impacts calculated for most of APSIM-Wheat parameters
revealed 42 parameters substantially impacting yield in most of the TPE.
Among those, a few parameters related to phenology, resource
acquisition, resource use efficiency and biomass allocation were
identified as potential candidates for crop improvement.\\While trait
variation was artificially set at \(\pm\) 20 \% and could be adjusted to
better mimic currently known genetic variability for traits of interest,
adjustments on the TPE could also be investigated. For instance, TPE for
future climate scenarios could be explored to identify potential traits
of future importance, providing crop models can properly deal with these
future conditions.\\To conclude, integrating GxExM interactions through
modeling approaches is an increasingly topical consideration to help
prioritizing investments of research efforts for the benefit of breeding
{[}17{]}. However, newly-gained computational knowledge has to be
constantly confronted to physiological reality in order to determine the
complexity of GxExM interactions that impede progress in crop
productivity.

\subsection{Supporting Information}\label{supporting-information}

\textbf{Figure S1. Range of variation used for function parameters.}
Each graph represents one function parameter (\emph{x} and \emph{y}
vectors), except for grouped parameters (i.e.~leaf, stem and pod
nitrogen demand). The graph titles match the \emph{Process} column in
table S1. Nominal values are indicted in green, while minimum and
maximum values are displayed blue and red, respectively. As some
parameters were grouped to be modified together, different symbols are
used for related processes (maximum, critical and minimum nitrogen
content) as defined in APSIM-wheat {[}29{]}.

\textbf{Table S1. Description of the APSIM-wheat parameters included in
the sensitivity analysis.} \emph{Module} refers to the sub-model where
the parameter is used in APSIM-wheat, \emph{Process} refers to the
physiological process targeted by the considered parameter and
\emph{Factor} is the parameter name used in the present study and in the
APSIM documentation {[}29{]}, where a complete description of the
parameters is given. The \emph{Default Value} field lists the nominal
value of the parameter for cultivar Hartog in APSIM-wheat 7.5 (only
first three values were presented when the parameter is defined as a
vector). In the \emph{Process} field, influential parameters in
indicated in bold and parameters that were grouped together for
physiologic reasons are identified by (*).

\newpage

\subsection*{References}\label{references}
\addcontentsline{toc}{subsection}{References}

1. Chapman S, Cooper M, Hammer G (2002) Using crop simulation to
generate genotype by environment interaction effects for sorghum in
water-limited environments. Australian Journal of Agricultural Research
53: 379--389.

2. Hammer G, Cooper M, Tardieu F, Welch S, Walsh B, et al. (2006) Models
for navigating biological complexity in breeding improved crop plants.
Trends in Plant Science 11: 587--593.

3. Chapman S (2008) Use of crop models to understand genotype by
environment interactions for drought in real-world and simulated plant
breeding trials. Euphytica 161: 195--208.

4. Chapman SC, Chakraborty S, Dreccer MF, Howden SM (2012) Plant
adaptation to climate change---opportunities and priorities in breeding.
Crop and Pasture Science 63: 251--268.

5. Zheng B, Chenu K, Fernanda Dreccer M, Chapman SC (2012) Breeding for
the future: What are the potential impacts of future frost and heat
events on sowing and flowering time requirements for australian bread
wheat (triticum aestivium) varieties? Global Change Biology 18:
2899--2914.

6. Chenu K, Cooper M, Hammer G, Mathews K, Dreccer M, et al. (2011)
Environment characterization as an aid to wheat improvement:
Interpreting genotype--environment interactions by modelling
water-deficit patterns in north-eastern australia. Journal of
Experimental Botany 62: 1743--1755.

7. Chenu K, Deihimfard R, Chapman SC (2013) Large-scale characterization
of drought pattern: A continent-wide modelling approach applied to the
australian wheatbelt--spatial and temporal trends. New Phytologist 198:
801--820.

8. Ortiz-Monasterio R, Sayre K, Rajaram S, McMahon M, others (1997)
Genetic progress in wheat yield and nitrogen use efficiency under four
nitrogen rates. Crop Science 37: 898--904.

9. Richards R, Hunt J, Kirkegaard J, Passioura J (2014) Yield
improvement and adaptation of wheat to water-limited environments in
australia---a case study. Crop and Pasture Science 65: 676--689.

10. Yang R-C, Blade SF, Crossa J, Stanton D, Bandara MS (2005)
Identifying isoyield environments for field pea production. Crop Science
45: 106--113.

11. Vega A de la, DeLacy I, Chapman S (2007) Changes in agronomic traits
of sunflower hybrids over 20 years of breeding in central argentina.
Field Crops Research 100: 73--81.

12. Chapman SC, Cooper M, Butler DG, Henzell RG (2000) Genotype by
environment interactions affecting grain sorghum. i. characteristics
that confound interpretation of hybrid yield. Aust J Agric Res 51:
197--208.

13. Cooper M, Woodruff DR, Eisemann RL, Brennan PS, DeLacy IH (1995) A
selection strategy to accommodate genotype-by-environment interaction
for grain yield of wheat: Managed-environments for selection among
genotypes. TAG Theoretical and Applied Genetics 90: 492--502.
doi:\href{http://dx.doi.org/http://dx.doi.org/10.1007/BF00221995}{http://dx.doi.org/10.1007/BF00221995}.

14. Mathews KL, Chapman SC, Trethowan R, Pfeiffer W, Van Ginkel M, et
al. (2007) Global adaptation patterns of australian and cIMMYT spring
bread wheat. Theoretical and Applied Genetics 115: 819--835.

15. Chapman SC, Crossa J, Edmeades GO (1997) Genotype by environment
effects and selection for drought tolerance in tropical maize. I. Two
mode pattern analysis of yield. Euphytica 95: 1--9.

16. Alwala S, Kwolek T, McPherson M, Pellow J, Meyer D (2010) A
comprehensive comparison between eberhart and russell joint regression
and gGE biplot analyses to identify stable and high yielding maize
hybrids. Field Crops Research 119: 225--230.

17. Hammer GL, McLean G, Chapman S, Zheng B, Doherty A, et al. (2014)
Crop design for specific adaptation in variable dryland production
environments. Crop and Pasture Science 65: 614--626.

18. Nyquist WE, Baker R (1991) Estimation of heritability and prediction
of selection response in plant populations. Critical reviews in plant
sciences 10: 235--322.

19. Cooper M, Stucker R, DeLacy I, Harch B (1997) Wheat breeding
nurseries, target environments, and indirect selection for grain yield.
Crop Science 37: 1168--1176.

20. Chapman S, Cooper M, Podlich D, Hammer G (2003) Evaluating Plant
Breeding Strategies by Simulating Gene Action and Dryland Environment
Effects. Agronomy Journal 95: 99--113.

21. Zheng B, Biddulph B, Li D, Kuchel H, Chapman S (2013) Quantification
of the effects of vRN1 and ppd-d1 to predict spring wheat (triticum
aestivum) heading time across diverse environments. Journal of
Experimental Botany 64: 3747--3761.

22. Chenu K, Chapman S, Tardieu F, McLean G, Welcker C, et al. (2009)
Simulating the yield impacts of organ-level quantitative trait loci
associated with drought response in maize: A`` gene-to-phenotype''
modeling approach. Genetics 183: 1507.

23. Jeuffroy M-H, Casadebaig P, Debaeke P, Loyce C, Meynard J-M (2014)
Agronomic model uses to predict cultivar performance in various
environments and cropping systems. a review. Agronomy for Sustainable
Development 34: 121--137.
doi:\href{http://dx.doi.org/10.1007/s13593-013-0170-9}{10.1007/s13593-013-0170-9}.

24. Rebetzke GJ, Chenu K, Biddulph B, Moeller C, Deery DM, et al. (2013)
A multisite managed environment facility for targeted trait and
germplasm phenotyping. Functional Plant Biology 40: 1--13.

25. Potgieter A, Hammer G, Butler D (2002) Spatial and temporal patterns
in australian wheat yield and their relationship with eNSO. Crop and
Pasture Science 53: 77--89.

26. Williams J, Hamblin AP, Hook RA (2002) Agro-ecological regions of
australia. methodologies for their derivation and key issues in resource
management. CSIRO Land; Water.

27. Keating BA, Carberry PS, Hammer GL, Probert ME, Robertson MJ, et al.
(2003) An overview of APSIM, a model designed for farming systems
simulation. European Journal of Agronomy 18: 267--288.

28. Holzworth DP, Huth NI, deVoil PG, Zurcher EJ, Herrmann NI, et al.
(2014) APSIM - evolution towards a new generation of agricultural
systems simulation. Environmental Modelling \& Software 62: 327--350.
doi:\href{http://dx.doi.org/http://dx.doi.org/10.1016/j.envsoft.2014.07.009}{http://dx.doi.org/10.1016/j.envsoft.2014.07.009}.

29. Zheng B, Chenu K, Doherty A, Chapman S (2014) The aPSIM-wheat module
(7.5 r3008). CSIRO. Available:
\url{http://www.apsim.info/Portals/0/Documentation/Crops/WheatDocumentation.pdf}.

30. Wang J, Wang E, Luo Q, Kirby M (2009) Modelling the sensitivity of
wheat growth and water balance to climate change in southeast australia.
Climatic Change 96: 79--96.

31. Saltelli A, Chan K (2000) Sensitivity Analysis. Scott EM, editor
Wiley.

32. Monod H, Naud C, Makowski D (2006) Working with dynamic crop models,
evaluation, analysis, parameterization and applications. In: Wallach D,
Makowski D, Jones JW, editors. Elsevier. pp. 55--100.

33. Valade A, Ciais P, Vuichard N, Viovy N, Caubel A, et al. (2014)
Modeling sugarcane yield with a process-based model from site to
continental scale: Uncertainties arising from model structure and
parameter values. Geoscientific Model Development 7: 1225--1245.
doi:\href{http://dx.doi.org/10.5194/gmd-7-1225-2014}{10.5194/gmd-7-1225-2014}.

34. Da Silva D, Han L, Faivre R, Costes E (2014) Influence of the
variation of geometrical and topological traits on light interception
efficiency of apple trees: Sensitivity analysis and metamodelling for
ideotype definition. Annals of botany 114: 739--752.

35. Martre P, He J, Le Gouis J, Semenov MA (2015) In silico system
analysis of physiological traits determining grain yield and protein
concentration for wheat as influenced by climate and crop management.
Journal of Experimental Botany: erv049.

36. Zhao G, Bryan BA, Song X (2014) Sensitivity and uncertainty analysis
of the aPSIM-wheat model: Interactions between cultivar, environmental,
and management parameters. Ecological Modelling 279: 1--11.

37. Morris MD (1991) Factorial sampling plans for preliminary
computational experiments. Technometrics 33: 161--174.

38. Campolongo F, Cariboni J, Saltelli A (2007) An effective screening
design for sensitivity analysis of large models. Environmental Modelling
\& Software 22: 1509--1518.

39. Wang E, Robertson M, Hammer G, Carberry P, Holzworth D, et al.
(2002) Development of a generic crop model template in the cropping
system model APSIM. European Journal of Agronomy 18: 121--140.

40. Zheng B, Holland E, Chapman S (2013) Wheat modelling: A case study
in innovating across cSIRO grid computing systems. eResearch Australasia
Brisbane, Australia.

41. Zomer RJ, Trabucco A, Bossio DA, Verchot LV (2008) Climate change
mitigation: A spatial analysis of global land suitability for clean
development mechanism afforestation and reforestation. Agriculture,
Ecosystems \& Environment 126: 67--80.

42. Iooss B, Lema{î}tre P (2014) A review on global sensitivity analysis
methods. arXiv preprint arXiv:14042405.

43. R Core Team (2014) R: A language and environment for statistical
computing. Vienna, Austria: R Foundation for Statistical Computing.
Available: \url{http://www.R-project.org/}.

44. Wickham H, Francois R (2015) Dplyr: A grammar of data manipulation.
Available: \url{http://CRAN.R-project.org/package=dplyr}.

45. Pujol G, Iooss B, Paul Lemaitre AJ with contributions from, Gilquin
L, Gratiet LL, et al. (2014) Sensitivity: Sensitivity analysis.
Available: \url{http://CRAN.R-project.org/package=sensitivity}.

46. Wickham H (2009) Ggplot2: Elegant graphics for data analysis.
Springer New York. Available: \url{http://had.co.nz/ggplot2/book}.

47. Poorter H, Evans JR (1998) Photosynthetic nitrogen-use efficiency of
species that differ inherently in specific leaf area. Oecologia 116:
26--37.

48. Poorter H, Garnier E (1999) Ecological significance of inherent
variation in relative growth rate and its components. Handbook of
functional plant ecology 20: 81--120.

49. Poorter H, Niinemets Ü, Poorter L, Wright IJ, Villar R (2009) Causes
and consequences of variation in leaf mass per area (lMA): A
meta-analysis. New Phytologist 182: 565--588.

50. Manschadi AM, Christopher J, Hammer GL, others (2006) The role of
root architectural traits in adaptation of wheat to water-limited
environments. Functional Plant Biology 33: 823--837.

51. Chapman SC, Hammer GL, Podlich DW, Cooper M (2002) Quantitative
genetics, genomics and plant breeding. In: Kang, editor. CAB
International, Wallingford UK. pp. 167--187.

52. Sadras V, Richards R (2014) Improvement of crop yield in dry
environments: Benchmarks, levels of organisation and the role of
nitrogen. Journal of Experimental Botany 65: 1981--1995.

53. Rebetzke GJ, Fischer RTA, Herwaarden AF van, Bonnett DG, Chenu K, et
al. (2014) Plot size matters: Interference from intergenotypic
competition in plant phenotyping studies. Functional Plant Biology 41:
107--118.

54. Bertin N, Martre P, Genard M, Quilot B, Salon C (2010) Under what
circumstances can process-based simulation models link genotype to
phenotype for complex traits? Case-study of fruit and grain quality
traits. Journal of Experimental Botany 61: 955--967.

55. Yin X, Struik PC (2010) Modelling the crop: From system dynamics to
systems biology. Journal of Experimental Botany 61: 2171--2183.

56. Veyradier M, Christopher J, Chenu K (2013) Quantifying the potential
yield benefit of root traits. In: Sievänen R, Nikinmaa E, Godin C,
Lintunen A, Nygren P, editors. Proceedings of the 7th international
conference on functional-structural plant models, saariselkä, finland,
9-14 june. iSBN 978-951-651-408-9. pp. 317--319.

57. Christopher J, Manschadi A, Hammer G, Borrell A (2008) Developmental
and physiological traits associated with high yield and stay-green
phenotype in wheat. Crop and Pasture Science 59: 354--364.

58. Snape J, Butterworth K, Whitechurch E, Worland A (2001) Waiting for
fine times: Genetics of flowering time in wheat. Euphytica 119:
185--190.

59. Reynolds M, Foulkes MJ, Slafer GA, Berry P, Parry MA, et al. (2009)
Raising yield potential in wheat. Journal of Experimental Botany 60:
1899--1918.

60. Bogard M, Ravel C, Paux E, Bordes J, Balfourier F, et al. (2014)
Predictions of heading date in bread wheat (triticum aestivum l.) using
qTL-based parameters of an ecophysiological model. Journal of
Experimental Botany 65: 5849--5865.

61. Richards RA (2000) Selectable traits to increase crop photosynthesis
and yield of grain crops. Journal of Experimental Botany 51: 447--458.

62. Parry MA, Reynolds M, Salvucci ME, Raines C, Andralojc PJ, et al.
(2010) Raising yield potential of wheat. iI. increasing photosynthetic
capacity and efficiency. Journal of Experimental Botany: erq304.

63. Zhu X-G, Long SP, Ort DR (2010) Improving photosynthetic efficiency
for greater yield. Annual review of plant biology 61: 235--261.

64. Reynolds M, Foulkes J, Furbank R, Griffiths S, King J, et al. (2012)
Achieving yield gains in wheat. Plant, Cell \& Environment 35:
1799--1823.

65. Fischer R (1996) Increasing yield potential in wheat: Breaking the
barriers. In: Reynolds M, Rajaram S, McNab A, editors. Workshop Proc.
Cd. Obregon, Mexico, CIMMYT. pp. 150--166.

66. Lopes M, Reynolds M, Manes Y, Singh R, Crossa J, et al. (2012)
Genetic yield gains and changes in associated traits of cIMMYT spring
bread wheat in a ``historic'' set representing 30 years of breeding.
Crop Science 52: 1123--1131.

67. Yang D, Luo Y, Ni Y, Yin Y, Yang W, et al. (2014) Effects of
exogenous aBA application on post-anthesis dry matter redistribution and
grain starch accumulation of winter wheat with different staygreen
characteristics. The Crop Journal 2: 144--153.

68. Christopher JT, Veyradier M, Borrell AK, Harvey G, Fletcher S, et
al. (2014) Phenotyping novel stay-green traits to capture genetic
variation in senescence dynamics. Functional Plant Biology 41:
1035--1048.

69. Babar M, Reynolds M, Van Ginkel M, Klatt A, Raun W, et al. (2006)
Spectral reflectance to estimate genetic variation for in-season
biomass, leaf chlorophyll, and canopy temperature in wheat. Crop Science
46: 1046--1057.

70. Pinto RS, Reynolds MP, Mathews KL, McIntyre CL, Olivares-Villegas
J-J, et al. (2010) Heat and drought adaptive qTL in a wheat population
designed to minimize confounding agronomic effects. Theoretical and
Applied Genetics 121: 1001--1021.

71. Rebetzke G, Condon A, Farquhar G, Appels R, Richards R (2008)
Quantitative trait loci for carbon isotope discrimination are repeatable
across environments and wheat mapping populations. Theoretical and
Applied Genetics 118: 123--137.

72. Dreccer MF, Herwaarden AF van, Chapman SC (2009) Grain number and
grain weight in wheat lines contrasting for stem water soluble
carbohydrate concentration. Field Crops Research 112: 43--54.

73. Abbad H, El Jaafari S, Bort J, Araus J (2004) Comparative
relationship of the flag leaf and the ear photosynthesis with the
biomass and grain yield of durum wheat under a range of water conditions
and different genotypes. Agronomie 24.

74. Tambussi EA, Bort J, Guiamet JJ, Nogu{é}s S, Araus JL (2007) The
photosynthetic role of ears in c3 cereals: Metabolism, water use
efficiency and contribution to grain yield. Critical Reviews in Plant
Sciences 26: 1--16.

75. Chenu K (2015) Characterising the crop environment--nature,
significance and applications. In: Sadras V, Calderini D, editors.
Academic Press. pp. 321--348.

76. Martre P, Quilot-Turion B, Luquet D, Memmah M-MO-S, Chenu K, et al.
(2015) Crop physiology: Applications for genetic improvement and
agronomy. In: Sadras VO, Calderini D, editors. Academic Press.
doi:\href{http://dx.doi.org/10.1016/B978-0-12-417104-6.00014-5}{10.1016/B978-0-12-417104-6.00014-5}.

77. Cabrera-Bosquet L, Molero G, Bort J, Nogu{é}s S, Araus J (2007) The
combined effect of constant water deficit and nitrogen supply on wUE,
nUE and \(\Delta\)13C in durum wheat potted plants. Annals of Applied
Biology 151: 277--289.

78. Sadras V, Rodriguez D (2010) Modelling the nitrogen-driven trade-off
between nitrogen utilisation efficiency and water use efficiency of
wheat in eastern australia. Field Crops Research 118: 297--305.

79. Sinclair TR, Purcell LC, Sneller CH (2004) Crop transformation and
the challenge to increase yield potential. Trends in Plant Science 9:
70--75.

80. Wallach D, Buis S, Lecharpentier P, Bourges J, Clastre P, et al.
(2011) A package of parameter estimation methods and implementation for
the sTICS crop-soil model. Environmental Modelling \& Software 26:
386--394.

81. Dumont B, Leemans V, Mansouri M, Bodson B, Destain J-P, et al.
(2014) Parameter identification of the sTICS crop model, using an
accelerated formal mCMC approach. Environmental Modelling \& Software
52: 121--135.

82. Messina CD, Podlich D, Dong Z, Samples M, Cooper M (2011)
Yield--trait performance landscapes: From theory to application in
breeding maize for drought tolerance. Journal of Experimental Botany 62:
855--868.

\end{document}